\documentclass[prd,preprint,aps,amsmath,amssymb,showpacs,superscriptaddress,nofootinbib]{revtex4-2}

\usepackage{xcolor}

\usepackage{graphicx} 
\usepackage{tabularx}
\usepackage{dcolumn}
\usepackage{bm}
\usepackage{hyperref}
\usepackage{color}
\usepackage[normalem]{ulem}
\usepackage{comment}
\usepackage{diagbox}
\usepackage{amsmath}
\renewcommand\[{\begin{equation}} 
\renewcommand\]{\end{equation}}
\usepackage{slashed}
\usepackage{mathtools}

\usepackage{subcaption} 
\usepackage{caption} 
\usepackage{float} 

\usepackage{ulem}
\renewcommand\[{\begin{equation}} 
\renewcommand\]{\end{equation}}

\begin{document}
\preprint{MI-HET-844}
\title{Imprints of Early Universe Cosmology on Gravitational Waves}

\author{James B. Dent}
\email{jbdent@shsu.edu}

\affiliation{Department of Physics, Sam Houston State University, Huntsville, TX, USA}

\author{Bhaskar Dutta}
\email{dutta@tamu.edu}

\affiliation{Mitchell Institute for Fundamental Physics and Astronomy,\\Department of Physics and Astronomy, Texas A\&M
University, College Station, USA}

\author{Mudit Rai}
\email{muditrai@tamu.edu}

\affiliation{Mitchell Institute for Fundamental Physics and Astronomy,\\Department of Physics and Astronomy, Texas A\&M
University, College Station, USA}

\date{\today}

\begin{abstract}
We explore the potential of gravitational waves (GWs) to probe the pre-BBN era of the early universe, focusing on the effects of energy injection. Specifically, we examine a hidden sector alongside the Standard Model that undergoes a strong first-order phase transition (FOPT), producing a GW signal. Once the phase transition has completed, energy injection initiates reheating in the hidden sector, which positions the hidden sector field so that additional phase transitions can occur. This can result in a total of three distinct phase transitions with a unique three-peak GW spectrum. Among these transitions, the first and third are of the standard type, while the intermediate second transition is inverted, moving from a broken to an unbroken phase. Using polynomial potentials as a framework, we derive analytical relations among the phase transition parameters and the resulting GW spectrum. Our results indicate that the second and third transitions generate GWs with higher amplitudes than the first, with a peak frequency ratio differing by up to an order of magnitude. This three-peak GW spectrum is detectable by upcoming facilities such as LISA, BBO, and UDECIGO. Notably, the phenomenon is robust across various potentials and model parameters, suggesting that hidden sector GWs provide a powerful tool for exploring new physics scenarios in the pre-BBN era.
\end{abstract}

\maketitle
\newpage

\section{Introduction}

With 90 candidate gravitational wave (GW) events since 2015 through observing run 3~\cite{KAGRA:2021vkt} of the LIGO Scientific, Virgo, and KAGRA Collaboration (LVK), along with recent evidence for the existence of nHz scale GWs from pulsar timing arrays ~\cite{NANOGrav:2023gor,EPTA:2023fyk,Reardon:2023gzh,Xu:2023wog}, GW astronomy has been established as a new window on the universe. With the advent of this field, exciting prospects for future studies across a wide range of frequencies with implications for astronomy, cosmology, and, most relevant to the current work, possible new physics scenarios have generated a great deal of excitement. One driver of such excitement is the possibility, due to GWs ability to propagate relatively freely from their production to the present, to probe interesting pre-Big Bang Nucleosynthesis (BBN) phenomena that are unreachable through other means.

The evolution of the universe from the time of BBN at a temperature of $T\sim\mathcal{O}(\textnormal{MeV})$ to the present  is relatively well understood through, for example, probes of light element abundances, the cosmic microwave background (CMB), and  large scale structure. Prior to BBN, and after an assumed inflationary phase, cosmology is less well established since Standard Model (SM) particles were tightly coupled in a thermal bath, negating the possibility of their producing observational relics from the pre-BBN universe. This is in contrast to GWs which, as they couple only gravitationally and are therefore not in equilibrium with the thermal bath, can provide information regarding whatever cosmic era they originate from, including those prior to BBN.

One may assume a solely radiation dominated $\Lambda$CDM universe from inflation to BBN, but the lack of observational evidence about this period allows for alternative possibilities. For example, a wide variety of 
scenarios such as the evaporation of primordial black holes (PBHs)~\cite{Hawking:1971ei, Carr:1974nx, Carr:1975qj,Kawasaki:2000en,Carr:2020gox, Villanueva-Domingo:2021spv}, the decay of moduli fields \cite{Moroi:1999zb,Dutta:2009uf},  
quantum fluctuations during inflation caused by a single field or multiple fields~\cite{Kodama:1984ziu, Mukhanov:1990me, Ma:1995ey, Lyth:1998xn}, and the collapse of cosmic string loops or domain walls \cite{Kibble:1976sj, Copeland:2009ga,Zeldovich:1974uw}, could exist where energy or entropy are injected locally in the early universe, leading to fluctuations in their density. Even with relatively small injections of energy, this local increase in energy can elevate the temperature of the surrounding medium. Imprints of such events happening in pre-BBN universe are not easy to track since the radiation domination remains unaltered. An interesting way to observe the effects of energy injection could be through first-order phase transitions \cite{Goldstone:1962es, Kirzhnits:1972iw, Coleman:1973jx,Dolan:1973qd,Kirzhnits:1976ts, Weinberg:1974hy, Witten:1984rs, hogan1986gravitational,Carrington:1991hz, Arnold:1992fb,Hindmarsh:2020hop, Domenech:2020ssp,Athron:2023xlk, Domenech:2024wao, Roshan:2024qnv} that induce gravitational waves via sound waves \cite{Kamionkowski:1993fg}, bubble collisions \cite{Kosowsky:1992rz, Kosowsky:1992vn}, and turbulence \cite{Kosowsky:2001xp}, potentially providing observable signatures.

In this work we explore the possible effects of energy injection in the context of a hidden sector that is thermally decoupled from the SM, with both sectors containing relativistic degrees of freedom (for a sampling of earlier studies on the cosmology of hidden sectors and phase transitions, see~\cite{Davoudiasl:2004be,Espinosa:2007qk,Espinosa:2008kw,Breitbach:2018ddu,Fairbairn:2019xog}). During the radiation era following inflation, the SM dominates the energy density of the universe.  If the hidden sector undergoes a phase transition, subsequent energy injections from decays can alter the thermal profiles of both sectors in such a way that multiple phase transitions occur. These multiple transitions could provide a striking observational signature in the form of a broad stochastic GW spectrum punctuated by a three peak structure. 
The envelope of such a broad GW spectrum could be observed simultaneously in multiple future detectors. The broadness of the peak (with a ratio of its frequency spread to peak frequency of $\Delta f/f \sim\mathcal{O}(10)$) can be a distinguishing feature for our scenarios which should be easily resolved based on the projected frequency resolutions of the GW experiments. 

The scenario envisioned is as follows (see Fig.~\ref{fig:Phases along hubble} for a schematic representation): a hidden sector field is sitting at the minimum of its potential, which evolves due to the temperature evolution of the universe. Eventually a first-order phase transition (phase 1) can occur as a lower minimum arises, provided there is a barrier between the false and true vacua. Now, however, after the first phase transition, a large enough energy injection can raise the temperature profile to where the field isn't immediately restored to the original minimum prior to the first transition, but rather situates the field so that it is in a false vacuum that then transitions back to the original minimum through a first-order phase transition (phase 2). Once it returns to the original minimum, it can undergo another first-order phase transition from the original vacuum as in the first instance (phase 3). Thus, we have three transitions providing three GW spectra.

For specificity, we choose polynomial-like potentials for the hidden sector \cite{Adams:1993zs,Dine:1992wr, Ellis:2019oqb}. However, the occurrence of multiple GW peaks due to multiple phase transitions is a general feature that can arise across a wide range of hidden sector energy scales and model classes. If the hidden sector either dominates or has comparable energy density to the SM, the Hubble parameter is significantly modified by the energy injection and any enhancements from later redshift transitions are suppressed or even erased, resulting in two gravitational wave peaks of similar amplitude.

The paper is organized as follows: In Section II, we focus on two specific reheating scenarios—moduli field decay and PBH reheating—and their effects on the hidden sector temperature. Section III discusses the various phase transitions and their differences. Section IV presents the resulting gravitational wave spectra and the main findings of our analysis. Conclusion and outlooks are presented in Section V. Appendix A contain explicit details on the reheating scenarios, focusing on PBH evaporation and moduli decay as two possible examples leading to energy injection. Appendix B covers details on phase transition parameters, and appendix C reviews the semi-analytical parameterization for GW spectrum.
\section{Energy injection in early universe}

In this section, we will present two scenarios of energy injection which lead to the reheating of both the hidden sector and the SM. When there is a large difference in energy densities between the two sectors, $\rho_h/\rho_{\textnormal{SM}}\ll 1$, we will see that the effect on the hidden sector is much greater than that on the SM, thus keeping the Hubble parameter largely unaffected since $H^2 \propto \rho_{\textnormal{tot}}$ where $\rho_{\textnormal{tot}} = \rho_{\textnormal{SM}}+\rho_{h}$. Note that many hidden sector models with light degrees of freedom that remain relativistic until or after BBN are constrained to have small energy budgets relative to the SM in order to evade $\Delta N_{eff}$ bounds~\cite{Planck:2018vyg}. In our setup, we consider cases where the light degrees of freedom go out of equilibrium before BBN, but in principle the alternative scenario can be easily accommodated.

This provides a window of opportunity to probe such energy injections via their impact on the phase transition history of the hidden sector. Depending on the details, we show that such injections can lead to three first order phase transitions, and hence can lead to three different gravitational wave peaks instead of one, as in the standard case. We find that these peaks are correlated and that the two new peaks have larger strain amplitudes than the first owing to them happening at a later time (or at a lower Hubble parameter value, since the SM is largely unaffected by smaller energy injections).

We want to point out that we focus on fast injections (compared to the Hubble rate), where the field remains effectively frozen as the bath temperature increases, experiencing a modified temperature-dependent potential. In the case of slower injections, the field follows its temperature-dependent minimum, retracing its initial expansion path. Both scenarios result in three transitions, differing in the second, inverse transition occurring immediately after injection.

We consider a hidden sector at a temperature $T$ that is thermally decoupled from the SM, and related to the SM temperature by $T = \xi_0T_{SM}$, where $\xi_0\leq1$ for our work. The total energy density is given by
\[
\rho_R(T) =\frac{\pi^2}{30} \left(g_h^* + \frac{g_{SM}^*}{\xi_0^4}\right)\,T^4
\]
where $g_h^*$ and $g_{SM}^* $ correspond to the degrees of freedom for the hidden sector and SM, respectively.
The energy injection implies that there is an effective rise in the thermal bath temperature, characterized by the parameter $\delta$, as 
\[
T\rightarrow \tilde{T} = T(1+\delta)
\]
For the SM to remain dominant over the hidden sector, there exists a simple condition on $\delta$ and $\xi_0$ given as,
\[
\label{eq:delta_condition0}
\xi_0(1+\delta)<\left(\frac{g^*_{SM}(T_i/\xi_0)}{g^*_h(T_i(1+\delta))}\right)^{1/4}
\]
where $T_i$ is the hidden sector temperature at the time of energy injection. Next, we provide the parameter $\delta$ in two example energy injection scenarios.

\subsection{Moduli decay to hidden sector}

First, we consider a scenario where a moduli field couples to the hidden sector. This leads to a preferential reheating of the hidden sector at temperatures where the moduli field falls out of equilibrium. Imposing energy conservation before and after the decay, we get (the details are shown in Appendix~\ref{app: moduli})
\begin{equation}
        \delta = \left(1+\frac{30\,m_\chi^2\chi_i^2}{\pi^2\,g_h^*\,T_h^4}\right)^{1/4}-1
\end{equation}
where $m_\chi$ is the mass of the moduli field and $\chi_i$ its corresponding initial field value after inflation. As shown explicitly in the Appendix \ref{app: moduli}, the $\delta$ values are governed by the choice of the aforementioned moduli field parameters. Also, the scale at which the fields inject the energy is correlated with the mass of the moduli field.

\subsection{PBH reheating}
Another instance where energy dumping in the early universe could occur is from PBH reheating via Hawking radiation. We emphasize that the impact of PBH reheating will fall more on the hidden sector than the SM due to its smaller energy content. Following the detailed arguments in Appendix \ref{app: PBH}, the temperature of the hidden sector receives a thermal kick given as
\begin{equation}
\delta = \left(1+\frac{\beta_{PBH}\,T_0}{T_h}\right)^{1/4} -1 
\end{equation}
where $\beta_{PBH}$ is the initial PBH mass fraction and $T_0$ corresponds to the PBH formation temperature. 

Having given an overview of the energy injection paradigm with two possible scenarios outlined, we will now see how such energy injection scenarios can provide multiple phase transitions, leading to unique GW signatures.

\section{Phase Transitions and Gravitational waves}

In this section, we focus on a standard polynomial potential for a hidden sector scalar field $\phi$ which receives thermal corrections in the early universe, where, depending on the model parameters, the potential allows for the possibility of a FOPT (refer to appendix A for more details about the model building). Note that the idea is not limited to the particular potential considered, which is chosen for the possibility of performing analytical calculations. Other models with FOPTs occurring at different energy scales than those considered here can also provide viable frameworks for multiple transitions.  Our potential is of the well-studied form,
\[
V(\phi,T) \approx D(T^2-T_0^2)\phi^2-E\,T\phi^3+\frac{\lambda}{4}\phi^4
\label{eq:V}
\]
where $D$, $E$, and $\lambda$ are model-dependent parameters.
At high temperature, $T\gg T_0$, $V(\phi,T)$ has a single minimum at $\phi = 0$. After the temperature drops below $T_1$ (defined in eq.(\ref{eq : T1}), there exists a second minimum, which is equipotential with the first at the critical temperature $T_c$, below which a phase transition occurs.

\begin{figure}[H]
\centering
\includegraphics[width=\textwidth]{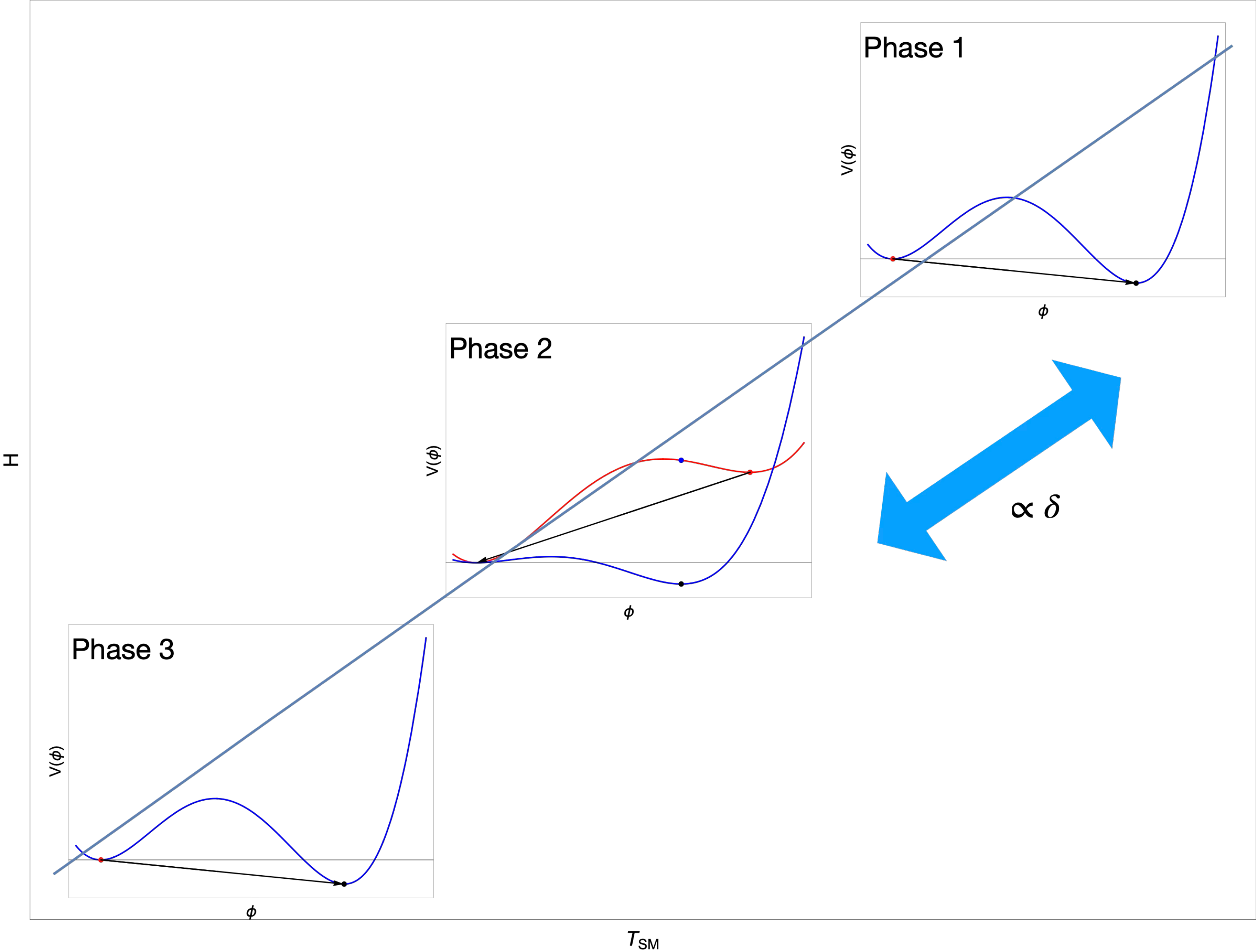}
\captionsetup{justification=justified}
\caption{Schematic view of the three phases along the Hubble rate as a function of temperature in the Standard Model (SM). The separation between the first transition and the second or third increases with larger injections, corresponding to larger $\delta$ values.}
\label{fig:Phases along hubble}
\end{figure}

Subsequent energy injection following the Phase 1 transition leads to a possibility of having three transitions, which we label as Phase 1, Phase 2 and Phase 3. The injection can occur at any time after the field is in the broken phase, say $\{\phi_i,T_i\}$, as shown in Fig. \ref{fig:Phases along hubble}.  For three transitions to occur, the following criteria for the energy injection must be met: 
\begin{align}
\label{eq:delta_conditions1}
& T_i< T_c<T_i(1+\delta)<T_1 \\
\label{eq:delta_conditions2}
& V(\phi_{min}(T_i(1+\delta))) > 0 \\ 
& \phi_{max}(T_i(1+\delta)) < \phi_i(T_i)
\end{align}
In terms of the model parameters, the above conditions lead to,
\[
\eta \coloneqq \frac{2\lambda\,D}{E^2} \in (9/4,9/2)
\label{eq: eta}
\]

\subsection{Phase 1}
\begin{figure}[H]
\centering
\includegraphics[width=0.75\linewidth]{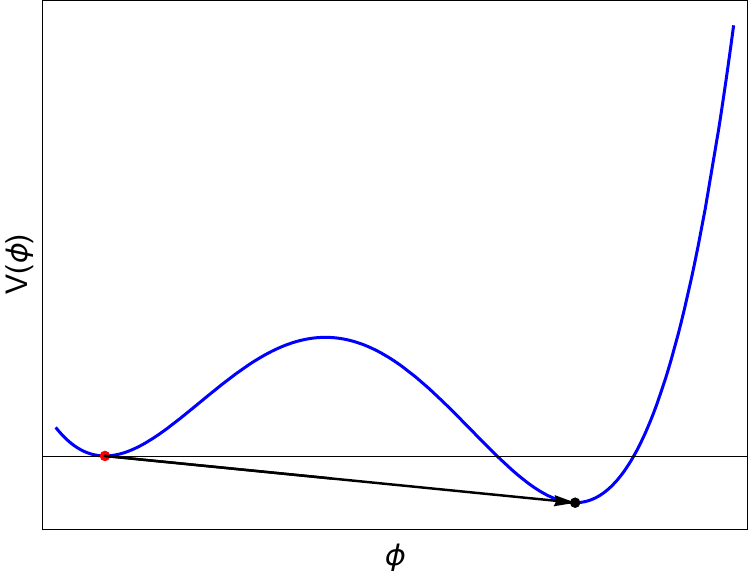} 
\caption{Depiction of a first order phase transition corresponding to a transition from a symmetric phase $\phi=0$ to a broken phase $\phi= \phi_c$.}
\label{fig:Phase I}
\end{figure}

The first phase is a classic first-order phase transition occurring as the universe cools. Initially, there is a global minimum at $\phi_0 = 0$, and as the universe cools, it eventually reaches the critical temperature $T_c$, where two equipotential minima appear at $\phi_0$ and $\phi_c$. As the temperature drops below $T_c$, bubbles of true vacuum nucleate and the field transitions from $\phi_0 \rightarrow \phi_c$, as shown in Fig.~\ref{fig:Phase I}. This transition is driven by the vacuum energy difference, while a frictional force on the bubble expansion arises as particles gain mass moving from the false vacuum (symmetric phase at $\phi_0$) to the true vacuum (broken phase). The transition is typically non-runaway, meaning the bubble wall remains sub-relativistic. We estimate the bubble wall velocity using Eq.(\ref{eq : wall_velocity}). We briefly recap the factors driving the phase transition. It results from a competition between the driving force from the latent heat of the vacuum and the frictional pressure from particles gaining mass as they cross from the false to the true vacuum. The driving force is defined by the latent heat which is given as  \footnote{In bag equation of state model \cite{Espinosa:2010hh}, the driving force or the $\alpha$ parameter is defined in terms of $\Delta V$, but we are  considering the latent heat to produce the driving force, as it is the net amount of energy difference between two vacua.} \cite{Athron:2023xlk},
\begin{equation}
    F_{dr}   \approx \alpha \, \rho_R  = \Delta \left(V - \frac{T}{4}\frac{dV}{dT}\right) 
\end{equation}
where $\Delta X= X_f -X_t $ corresponds to the difference of quantities between false vacua and the true vacua. For our potential this expression becomes
\begin{align}
    \Delta \left(V - \frac{T}{4}\frac{dV}{dT}\right)  
     &= -\frac{\phi_c^2}{4}(2D(T_N^2-2\,T_0^2)-3\,E\,T_N\,\phi_c + \lambda \phi_c^2) \\ \nonumber
     & = \frac{T_N^2\,\phi_c^2}{2\lambda} \left( \lambda \, D\frac{T_N}{T_c} - E^2 \right) > 0 
\end{align}
where $T_N$ is the nucleation temperature  defined in Appendix~\ref{app:parameters} and the inequality in the second line imposes a positivity condition on the latent heat. The above quantity will act as the driving force for the usual phase transitions, and is positive as long as the nucleation temperature is not much smaller than the critical temperature (since $\lambda D > E^2$ from phase transition conditions discussed in Appendix~\ref{app:parameters}).
In the case of runaway scenarios, the boost factor of the bubble wall, $\gamma_w \rightarrow \infty $, and the friction term due to particle mass gain, $\Delta p_{LO}$ is roughly given by \cite{ Espinosa:2010hh, Ellis:2019oqb,Barni:2024lkj},
\begin{equation}
\Delta p_{LO}^{\gamma_w \rightarrow \infty} \approx \sum_i \frac{c_i\,g_i \, (m_{t,i}^2-m_{f,i}^2)}{24} > 0
\end{equation}
where the constants $c_i$ are $c_i = 1 $ for bosons and $c_i =1/2$ for fermions, $m_{t,i}$ and $m_{f,i}$ are the masses in the true and false phases respectively, and $g_i$ is the number of degrees of freedom. The regime $F_{dr} > \Delta p_{LO}^{\gamma_w \rightarrow \infty}$ corresponds to runaway transitions.
\subsection{Phase 2}
\begin{figure}[H]
\centering
\includegraphics[width=0.85\linewidth]{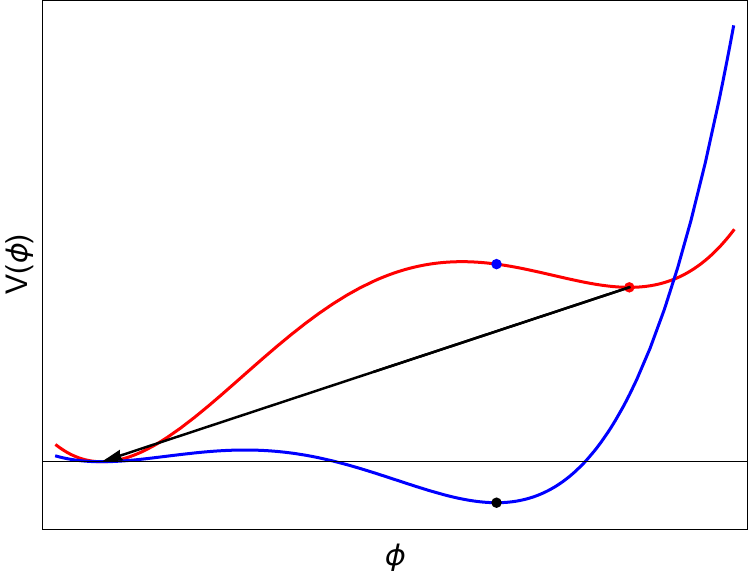} 
\caption{A first-order phase transition corresponding to the field going from the broken phase $\phi =\phi_i$ to the symmetric phase $\phi = 0$, where the blue curve represents the state before energy injection and the red curve represents the state after energy injection, assuming the injection rate is faster than the Hubble rate.} 
\label{fig:Phase II}
\end{figure}

Consider a scenario where an energy injection occurs after the initial phase transition, raising the hidden sector temperature from $T_i \rightarrow T_i(1 + \delta)$ while the field remains at $\phi_i$\footnote{We are assuming the rapid energy injection scenario where the injection rate exceeds the Hubble parameter.}. The field then rolls to its new, temperature-dependent minimum, ${\phi(T_i(1+\delta)), T_i(1+\delta)}$, eventually transitioning to $\phi=0$, which has a lower potential. We label this as Phase 2, shown in Fig. \ref{fig:Phase II}.

This transition is unique because the friction force is negative: particles lose mass as they move from the false vacuum (broken phase) to the true vacuum (symmetric phase). The driving force, estimated from the latent heat, is also negative and acts in opposition\footnote{\cite{Barni:2024lkj} discusses inverse transitions within the bag model framework, noting that the opposing force is counterbalanced by mass loss and shifts in relativistic degrees of freedom as particles enter the symmetric phase.}. Thus, the net driving force for this transition primarily arises from particle mass loss at leading order, which is given as,
\begin{align}
    F_{opp}  & \approx \alpha \, \rho_R  = \Delta \left(V - \frac{T}{4}\frac{dV}{dT}\right) \\ \nonumber 
     &= \frac{\phi_i^2}{4}(2D(T_c^2-2\,T_0^2)-3\,E\,T_c\,\phi_i + \lambda \phi_i^2) \\ 
     \implies  F_{opp}&\approx -T_i(1+\delta)\,\eta \,(4\,D\,T_i(1+\delta) - 3\,E\, \phi_i) -2\,D\,T_0^2 < 0 \nonumber
\end{align}
up to $O(\eta)$, where $T_c = T_i(1+\delta) (1- \eta)$\footnote{$F_{opp}<0$ holds as long as $T_N$ is not much smaller than $T_c$}.
The driving force is sourced from the leading pressure term which, in the runaway scenario, can be defined as \cite{Barni:2024lkj},
\[
\Delta p_{LO}^{\gamma_w \rightarrow \infty} \approx \sum_i \frac{c_i\,g_i \, (m_{t,i}^2-m_{f,i}^2)}{24} < 0
\]
A runaway transition occurs if $|\Delta p_{LO}^{\gamma_w \rightarrow \infty}| > |F_{opp}|$ \cite{Barni:2024lkj}. In both the runaway\footnote{Recent studies indicate that in models with vector particles and gauge symmetries, NLO friction contributions are positive, leading to $\gamma_w \rightarrow \gamma_{eq}$ and preventing a runaway phase \cite{Azatov:2024auq}.} and non-runaway scenarios, the driving force originates from particle mass loss, while the negative latent heat deters the phase transition.
 
\subsection{Phase 3}
After the second phase transition,  the universe continues to cool and another phase transition occurs from the symmetric phase at $\phi = 0$ to the broken phase $\phi = \phi_c$, similar to that of the initial phase 1 transition, as depicted in Fig.~\ref{fig:Phase I}. We refer to this as phase 3, which happens at a lower Hubble value as compared to phase 1 and phase 2. 

In the following sections, we will explore the three phase transitions in detail, comparing their relative strengths and examining how cosmology influences the correlations among various phase transition parameters.

\subsection{Gravitational waves}
Gravitational waves are produced from the first order phase transitions occurring in the hidden sector through three main mechanisms: bubble collisions, sound waves, and magneto-hydrodynamic (MHD) turbulence. Regardless of the source, the relevant
physical quantity characterizing the stochastic GW signal is the differential GW density parameter\cite{Huber:2008hg,Caprini:2009yp},
\begin{align}\label{eq:spectrum}
  \Omega_\text{GW}(f) \equiv
  \frac{1}{\rho_c} \frac{d\rho_\text{GW}(f)}{d\log f} 
\end{align}
where $\rho_c = 3 H^2 / (8 \pi G_N)$ is the critical energy density. To derive the observed spectrum today, it is necessary to consider the expansion of the universe from the moment the GWs were emitted to the present. This expansion causes redshift of the energy density and the frequency of the gravitational waves~\cite{Breitbach:2018ddu},
\begin{align}
  \label{eq:full-redshift}
  \Omega_\text{GW}^0(f) &= \mathcal{R}\,\Omega_\text{GW}\left(\frac{a_0}{a}f\right)
\end{align}
where
\begin{equation}
\mathcal{R} \equiv \left(\frac{a}{a_0}\right)^4 \left(\frac{H}{H_0}\right)^2 \simeq 2.473 \times 10^{-5} \, h^{-2} \left( \frac{g_s^\text{EQ}}{g_s} \right)^{4/3} \left(\frac{g_\rho}{2}\right)
\end{equation}
Here $\Omega_\text{GW}^0(f) \,(\Omega_\text{GW}) $ denotes the spectrum today (at emission), and $a_0\,(a)$ is the scale factor today (at nucleation).

The functional characterization of the GW spectrum depends on the dominant production mechanism, which in turn is dependent on the class of the transition. In runaway transitions, the bubble walls often become ultra-relativistic \cite{Breitbach:2018ddu}, with bubble collisions and sound waves being the primary sources of GW emission. In non-runaway cases, sound waves dominate GW production, with turbulence contributing as well. To address these scenarios for first-order phase transitions, we use a detailed semi-analytical approach outlined in Appendix~\ref{app:GW}. Although we employ the latest results in the literature, uncertainties still remain, and improvements are anticipated as research advances in this area.

The key features of GWs produced by phase transitions are the peak amplitude $\Omega^{(p)}_{\rm GW}$, and the frequency $f_p$ at which the peak amplitude occurs. In the usual case of a single phase transition, these are determined by the thermal parameters calculated from the underlying particle physics model. In the case we are considering, where multiple phase transitions occur, the energy injection plays an important role in the determination of $\Omega^{(p)}_{\rm GW}$ and $f_p$ for the two subsequent transitions, in addition to the usual thermal parameters.

\subsection{Relationship amongst the phase parameters}
\subsubsection{Relation between phase 2 and phase 3}
We can relate the phase transition parameters - the action and the rate and the strength of the phase transition for phase 2 - in terms of phase 3, specifically at their respective nucleation temperatures. This provides estimates on the possible hierarchy between the different phase transition parameters for polynomial-like potentials, without having to rely on scanning the individual model parameters. Following \cite{Ellis:2020awk, Adams:1993zs}, we can write an analytical expression for the Euclidean action as follows,
\[
\frac{S_3}{T} \approx \frac{8\,E}{\lambda^{3/2}}f_S(\kappa(T))
\label{eq:S3overT}
\]
where 
\[
f_S(x) = \frac{\left(8 \pi  \sqrt{x}\right) \left(0.818 x^3-5.533 x^2+8.2938 x\right)}{81 (2-x)^2}
\]
and $\kappa(T)$ is defined in Eq.(\ref{eq: kappa}). Now using the equations~(\ref {eq: Phase_II_E}) and (\ref{eq: Phase_II_kappa}), we get
\[
\left(\frac{S_3}{T}\right)_2 \approx \left(\frac{S_3}{T}\right)_3\,\frac{f_S(\tilde{\kappa}(T,\tilde{T}_i))}{f_S(\kappa(T))}\left(\frac{1+\sqrt{9-4\kappa(\tilde{T}_i)}}{2}\right)
\]
The condition for a phase transition to occur leads to\footnote{here $j \in\{1,3\}$ corresponds to phase 1 and phase 3 and $\kappa_j = \kappa(T_{N,j})$},
\[
\left(\frac{S_3}{T}\right)_j \approx 173.7-2\log g_{*,j} + 8\log \xi_j -4\log \frac{T_0}{{\rm GeV}} + 2 \log \left(1-\frac{\kappa_{N,j}}{\eta}\right)
\label{eq : S3hubble}
\]
Using the approximate definition for the respective nucleation temperatures, we can recast the above expressions in order to extract the ratios of the $f_S$ functions, which will yield the allowed range of $\kappa(T_{N,2})$ as a function of $\kappa(T_{N,3})$ and $\kappa(\tilde{T_i})$. The previous equation yields,
\[
\frac{f_S(\tilde{\kappa}(T_{N,2},\tilde{T}_i))}{f_S(\kappa_{N,3})} \approx 1-\frac{2\log \left(\frac{\eta - \kappa_{N,2}}{\eta-\kappa_{N,3}}\right)}{173.7-2\log g_{*,3} + 8\log \xi_f -4\log \frac{T_0}{{\rm GeV}} + 2 \log \left(1-\frac{\kappa_{N,3}}{\eta}\right)}
\]
where $T_{N,x} = \frac{T_0}{\sqrt{1-\frac{\kappa_{N,x}}{\eta}}},\,x\in\{1,2,3\}$.  
The explicit model dependence is only logarithmic, which enables us to place robust bounds on the ratios of various phase transition parameters between phase 2 and phase 3. Note that we always require the relations $T_c<\tilde{T_i}<T_1$ and $T_0<T_{N,j}<T_c$, which implies that the allowed $\kappa$ values are given as 
\begin{equation}
         0< \kappa(T_{N,j}),\,\tilde{\kappa}(\tilde{T}_i,T_{N,2})< 2, \quad
    2 < \kappa(\tilde{T}_{N,2})<\kappa(\tilde{T}_i) < 9/4
\end{equation}

The peak GW frequency is proportional to $\beta$, which characterizes the rate of the phase transition through the ratio $\beta/H$, determined by the model parameters $\eta$, $T_0$, and $\xi$. We can obtain the analytical formula for the transition rate from the above action \cite{Ellis:2020awk} for phase 1 and phase 3 as,
\begin{equation}
    \left(\frac{\beta}{H}\right)_j = 24\pi\,\frac{8\,D}{\sqrt{\lambda}\,E}\frac{T_0^2}{T_{N,j}^2}\,f_\beta(\kappa_{N,j}),\quad j=\{1,3\}
\end{equation}
From this expression we find that phase 1 and phase 3 have similar $\beta/H$ values. For phase 2, the analogous expression is,
\[
\left(\frac{\beta}{H}\right)_2 = 24\pi\,\frac{8\,\tilde{D}}{\sqrt{\lambda}\,\tilde{E}}\frac{T_0^2}{T_{N,2}^2}\,f_\beta(\tilde{\kappa}(\tilde{T}_i,T_{N,2}))
\]
where,
\begin{equation}
    f_\beta(x) = \frac{2 \sqrt{x} \left(0.818 (3 x-14) x^2-5.533 (x-10) x-8.2938 (x+6)\right)}{243 (x-2)^3} = \frac{ f_S'(x)}{6\,\pi}
\end{equation}
with $\tilde{D}$ and $\tilde{E}$ defined in Eqs.(\ref{eq: Phase_II_D}) and~(\ref{eq: Phase_II_E}).
The ratio between the transition rates is then given as,
\[
\frac{(\beta/H)_2}{(\beta/H)_3} = \left(\frac{\eta - \kappa_{N,2}}{\eta-\kappa_{N,3}}\right)\,\frac{\sqrt{9-4\,\kappa_i}}{\,\kappa_i}\left(\frac{\sqrt{9-4\,\kappa_i}+3}{\sqrt{9-4\,\kappa_i}+1}\right) \frac{f_\beta(\tilde{\kappa}(\tilde{T}_i,T_{N,2}))}{f_\beta(\kappa_{N,3})}
\label{eq : beta_ratio}
\]
where the first term encodes explicit model information via $\eta$ and $\left(\frac{T_{N,3}}{T_{N,2}}\right)^2 = \left(\frac{\eta - \kappa(T_{N,2})}{\eta-\kappa(T_{N,3})}\right)<1$.
Following Eq.(\ref{eq : beta_ratio}), phase 2 has a smaller $\beta/H$ value, causing the peak frequency for phase 2 to shift towards lower values. 

The peak frequencies \footnote{modulo the wall velocity factors, which for sounds waves would lead to an additional $v_{w,3}/v_{w,2}$ factor that increases the peak separation} are scaled by the respective Hubble values. This removes explicit model dependence as $\xi_f$ is the same for both phase 2 and phase 3, and the overall Hubble term takes care of the $T_N^2$ factors. This yields,
\begin{equation}
  \frac{f_{p,2,em}}{f_{p,3,em}}  \approx \frac{\beta_2}{\beta_3} \approx   \frac{\sqrt{9-4\,\kappa_i}}{\kappa_i}\left(\frac{\sqrt{9-4\,\kappa_i}+3}{\sqrt{9-4\,\kappa_i}+1}\right) \frac{f_\beta(\tilde{\kappa}(\tilde{T}_i,T_{N,2}))}{f_\beta(\kappa_{N,3})}
  \label{eq: emitted_freq_ratio}
\end{equation}
and at the time of observation we get,
\begin{equation}
  \frac{f_{p,2,0}}{f_{p,3,0}}  \approx   \sqrt{\frac{\eta - \kappa_{N,3}}{\eta-\kappa_{N,2}}}\,\frac{\sqrt{9-4\,\kappa_i}}{\kappa_i}\left(\frac{\sqrt{9-4\,\kappa_i}+3}{\sqrt{9-4\,\kappa_i}+1}\right) \frac{f_\beta(\tilde{\kappa}(\tilde{T}_i,T_{N,2}))}{f_\beta(\kappa_{N,3})}
\end{equation}

The strength of the phase transition scales are given by $\alpha \propto \xi^4 \alpha_h$, where $\alpha_h$ represents the strength of the phase transition relative to the relativistic energy of the hidden sector. We can express $\alpha_h$ for phase 1 and phase 3 as \cite{Ellis:2020awk},
\begin{equation}
    \alpha_{h,j} \approx \frac{128\,E^2\,D}{\lambda^2\,g_h} \frac{T_0^2}{T_{N,j}^2}\,f_\alpha(\kappa_{N,j})
\end{equation}
and for phase 2 as,
\begin{equation}
    \alpha_{h,2} \approx \frac{128\,\tilde{E}^2\,\tilde{D}}{\lambda^2\,g_h} \frac{T_0^2}{T_{N,2}^2}\,f_\alpha(\tilde{\kappa}(T_i,T_{N,2}))
\end{equation}
where 
\begin{equation}
    f_\alpha(x) = \frac{15 \left(-2 x+3 \sqrt{9-4 x}+9\right)}{256 \pi ^2}
\end{equation}
This yields the ratio of the strength parameters for phase 2 and phase 3,
\begin{equation}
    \frac{\alpha_{h,2}}{\alpha_{h,3}} \approx \left(\frac{\eta - \kappa_{N,2}}{\eta-\kappa_{N,3}}\right)\,\frac{\sqrt{9-4\,\kappa_i}}{8\kappa_i}\,\left((\sqrt{9-4\,\kappa_i}+3)(\sqrt{9-4\,\kappa_i}+1)\right)^2\,\frac{f_\alpha(\tilde{\kappa}(T_i,T_{N,2}))}{f_\alpha(\kappa_{N,3})}
    \label{eq: alpha_ratio}
\end{equation}

We can express $\eta$ in terms of phase transition parameters as,
\begin{align}
     \eta &  = \kappa_j + \frac{1}{12 \pi}\,\frac{(\beta/H)_j}{f_\beta(\kappa_j)}\,\frac{f_S(\kappa_j)}{(S_3/T)_j} 
    \label{eq : eta_para} \\ 
    & = \kappa_2 + \frac{\left(\beta/H\right)_2}{2\,\left(S_3/T\right)_2}\,\frac{1}{(\log(f_S(\tilde{\kappa}(\kappa_2,\kappa_i))))'\,f_\kappa(\kappa_i)}
    \label{eq : eta_para2}
\end{align}
By combining the above expressions with the bounds on $\eta$, we obtain a range of allowed values for $\beta/H$ as a function of $S_3/T$. This varies logarithmically with $\xi$ and $T_0$, and is typically $\mathcal{O}(1000)$, with $\kappa \in (0,2)$. Once the combinations $\eta - \kappa_j$ and $\eta - \kappa_2$ from Eqs.(\ref{eq : eta_para}) and~(\ref{eq : eta_para2}) are substituted into Eqs.(\ref{eq : beta_ratio}) and~(\ref{eq: alpha_ratio}), the ratios of $\beta/H$ (and thus $f_p$) and $\alpha_h$ between phase 2 and phase 3 decrease monotonically with increasing $\eta$. This means that maximizing the peak frequency separation results in greater suppression of peak amplitudes in phase 2 relative to phase 3. 

\subsubsection{Relation between phase 1 and phase 3}
Phase 1 and phase 3 transitions are similar in nature, differing mainly in the redshift values at which they occur. This difference is primarily driven by $\delta$, which quantifies energy injection. The conditions in Eqs.(\ref{eq:delta_condition0}), (\ref{eq:delta_conditions1}), and (\ref{eq:delta_conditions2}) ensure that $\delta$ is bounded, preventing $\kappa_3$ from differing significantly from $\kappa_1$. As $\delta$ increases, the temporal separation between phases 2 and 3 relative to phase 1 grows, leading to greater peak separation. Conversely, a decrease in $\delta$ reduces the separation between phases 1 and 3, causing their spectra to converge due to minimal redshift effects. To quantify the above discussion, consider the relation,
\[
\kappa_3 \approx \kappa_1 + \chi,\quad \chi \ll 1 
\]
where $\chi$ parameterizes the separation between $\kappa_1$ and $\kappa_3$. Following Eq.(\ref{eq : S3hubble}) we get,
\[
\left(\frac{S_3}{T}\right)_3 \approx \left(\frac{S_3}{T}\right)_1 + 2 \log \left(\frac{\eta-\kappa_{3}}{\eta-\kappa_1}\right) + 8\log(1+\delta)
\]
Another relation between the two quantities comes from Eq.(\ref{eq:S3overT}), yielding,
\[
\left(\frac{S_3}{T}\right)_3 \approx \left(\frac{S_3}{T}\right)_1 \frac{f_S(\kappa_3)}{f_S(\kappa_1)}
\]
We can simplify the ratios of the $f_S$ functions in terms of $f_\beta$ since, 
\begin{align}
  & f_S(\kappa_3) \approx f_S(\kappa_1) + \chi\,f_S'(\kappa_1) \\ \nonumber
  \implies & \frac{f_S(\kappa_3)}{f_S(\kappa_1)} \approx 1+\frac{\chi\,6\pi\,f_\beta(\kappa_1)}{f_S(\kappa_1)}
\end{align}
and using Eq.(\ref{eq : eta_para}), we get, 
\[
\frac{6\,\pi\,f_\beta(\kappa_i)\,(S_3/T)_i}{f_S(\kappa_i)} = \frac{\left(\beta/H\right)_i}{2(\eta -\kappa_i)}
\]
Combining the preceding equations we get,
\[
\left(\frac{S_3}{T}\right)_3 \approx \left(\frac{S_3}{T}\right)_1 +\frac{\chi\,(\beta/H)_1}{2\,(\eta-\kappa_1)}
\]
Finally, we can use the above equation and Eq.(\ref{eq : S3hubble}) to arrive at,
\begin{align}
    \frac{\chi\,(\beta/H)_1}{2\,(\eta-\kappa_1)}  & \approx 8\log(1+\delta) + 2\log(1-\frac{\chi}{\eta-\kappa_1}) \\ 
    \implies  \kappa_3 &\approx \kappa_1 + \frac{16\,\log(1+\delta)\,(\eta-\kappa_1)}{(\beta/H)_1}
   \\
     & \approx  \kappa_1 +  \frac{8\,\log(1+\delta)}{(S_3/T)_1\,(\log(f_S(\kappa_1)))'}\, \label{eq: kappa_3_1}
\end{align}

Thus, we can extract $\kappa_3$ from $\kappa_1$ based on the model, the scale of the potential, and the energy injection. The other cosmological parameters, namely $T_0$ and $\xi$, contribute only logarithmically with a support of $\mathcal{O}(100)$ factor. Thus, slight changes in these parameters would not significantly alter the overall behavior. Since the thermal parameters and resulting gravitational wave spectrum for phase 3 directly depends on $\kappa_3$, there is a clear relationship between the GW signals from the two phase transitions, providing insights into early universe energy injection.

From the above discussion with $\kappa_3\approx \kappa_1$, it is straightforward to infer that the phase transition strength parameter ratio between phase 1 and phase 3 would be close to 1. We get, 
\[
\frac{\alpha_{h,3}}{\alpha_{h,1}} \approx 1+\frac{8\,\log(1+\delta)}{(S_3/T)_1}\left(\frac{(\log(f_\alpha(\kappa_1))'}{(\log(f_S(\kappa_1))'}-\frac{(S_3/T)_1}{(\beta/H)_1}\right)
\label{eq: ah3_ah1}
\]
Each of the ratios in the difference within the parentheses in the second term is $\lesssim\mathcal{O}(1)$, and the outside factor is $\lesssim\mathcal{O}(1/10)$, which gives the overall ratio $\alpha_{h,3}/\alpha_{h,1}\approx 1$.

Similarly, the peak frequencies also differ by small amounts given as,
\begin{equation}
        \frac{f_{p,obs,3}}{f_{p,obs,1}} \approx \sqrt{\frac{\eta - \kappa_{N,1}}{\eta-\kappa_{N,3}}}\,\frac{f_\beta(\kappa_3)}{f_\beta(\kappa_1)}\approx 1+ \frac{8\,\log(1+\delta)}{(S_3/T)_1}\left(\frac{(S_3/T)_1}{(\beta/H)_1}
+ \frac{\log(f_\beta(\kappa_1))'}{\log(f_S(\kappa_1))'}\right)
\label{eq: fp3_fp1}
\end{equation}
which leads to $f_{p,obs,3}/f_{p,obs,1}\approx 1$.

The GW amplitude scales inversely with powers of $\beta/H$, hence lower values lead to stronger peak amplitudes. As the strength parameter $\alpha_h$ increases, the amplitude of the gravitational waves also increases. For fixed cosmological parameters ($T_0$, $\delta$, $\xi_{0/f}$) and fixed $\kappa$, larger $\eta$ leads to a lower nucleation temperature, implying less redshift suppression GW amplitudes for all the three phases.

Lastly, we would like to point out that the gravitational wave spectrum is also influenced by the wall velocity, $v_w$ — the amplitude decreases while the peak frequency increases with a lower $v_w$. For our calculations, we adopt the semi-analytic form in Eq.(\ref{eq : wall_velocity}). However, if $v_w$ were treated as an adjustable external parameter, then for non-runaway transitions (phases 1 and 3), a smaller $v_w$ would lead to lower peak amplitudes and higher peak frequencies, overall producing a broader GW spectrum.

\subsection{Impact of cosmology on GW spectrum}

In this subsection, we aim to clarify the impact of cosmological parameters on the peak frequencies and gravitational wave amplitudes, using the semi-analytical parameterization outlined in Appendix \ref{app:GW}.

The parameters $\xi_0$ and $\xi_f$ represent the temperature ratios between the hidden sector and the Standard Model before and after the kick from energy injection, primarily influencing the GW amplitude and frequency through the explicit redshift factors. Implicitly, the nucleation temperature depends logarithmically on $\xi_{0/f}$, leading to a small effect. As $\xi_{0/f}$ decreases, the transition occurs at earlier times (larger Hubble values), thereby reducing the gravitational wave amplitudes (due to a greater redshift suppression of $\alpha$), and a shift of the peak frequency to higher values. To make it more explicit, we can express the total phase transition strength parameter $\alpha$ in terms of the hidden sector transition strength parameter $\alpha_h$ given via,
\[
\alpha_1 = \alpha_{h,1} \,\xi_0^4,\,\quad  \alpha_{2} =\alpha_{h,2}\, \xi_f^4 ,\,\quad \alpha_{3} =\alpha_{h,3} \,\xi_f^4 
\label{eq: alpha_xi}\] where $\xi_f = \xi_0(1+\delta)$. Thus, the dominant contribution of the redshift effect is encoded in terms of powers of $\xi_{0/f}$ explicitly for GW amplitude. 

To extract $\xi_{0/f}$, we assume a GW spectrum is observed, meaning we have knowledge of one peak amplitude and frequency. Without loss of generality we can start by assuming these values correspond to those produced by the phase 3 transition (if one assumed this was from phase 2, then the following procedure could be done in reverse; phase 1 is typically sub-dominant to phase 3, and thus less likely to be observed.). From this, we can derive model parameters using the phase transition relations outlined earlier. The second phase transition can occur within the range $2 < \kappa_i < 2.25$, but we focus on maximizing peak separation, allowing us to fix $\kappa_i$, $\kappa_2$, and $\kappa_3$. From Eq.(\ref{eq: emitted_freq_ratio}), we can calculate the emitted peak frequency for phase 2 from the (observed) emitted peak frequency of phase 3. We can then solve for $\xi_f$ as,
\begin{align}
   \xi_f = & \frac{\gamma\,T_0^2}{M_\text{pl}} \left( \frac{192\, \pi  \, D}{\sqrt{\lambda} \, E \, (\beta_2 - \beta_3)} \left( f_\beta(\tilde{\kappa}) \left( \sqrt{9 + 4\kappa_i} \left( 1 + \frac{2}{1 + \sqrt{9 - 4\kappa_i}} \right) \right) - f_\beta(\kappa_3) \right) \right)^{1/2} \\ \nonumber
   \approx & \frac{\gamma\,T_0^2}{M_\text{pl}} \left( \frac{192\, \pi \, D}{\sqrt{\lambda} \, E \, (f_{p,em2}/v_{w2} - f_{p,em3}/v_{w3})} \left( f_\beta(\tilde{\kappa}) \left( \sqrt{9 + 4\kappa_i} \left( 1 + \frac{2}{1 + \sqrt{9 - 4\kappa_i}} \right) \right) - f_\beta(\kappa_3) \right) \right)^{1/2}
\label{eq: xifromphase23} 
\end{align}
Since $f_{p,\,em,\,i} \propto \beta_i$, knowing the peak frequency difference directly yields the value of $\xi_f$ for the hidden sector, in terms of model parameters and the scale of the hidden sector $T_0$ up to the factor of bubble wall velocity.

We can also compare the GW peak amplitudes for phase 1 and phase 3 and relate it to the energy injection. It is reasonable to focus on the scenario where both of the produced GW are sourced by sound waves. Then following the semi-analytical ansatz detailed in the appendix as well as the relations derived in the previous subsections, we get,
\[
\frac{\Omega_{{\rm GW},3,obs}^{(p)}}{\Omega_{{\rm GW},1,obs}^{(p)}} \approx (1+\delta)^6\,\left(\frac{(\beta/H)_3}{(\beta/H)_1}\right)^{\rm{eff}-1/2}\,\left(\frac{f_\beta(\kappa_1)\,f_\alpha(\kappa_3)}{f_\beta(\kappa_3)\,f_\alpha(\kappa_1)}\right)^{\rm{eff}+3/2}
\label{eq:omega31ratio}
\]
where the `eff' factor in the exponents denotes the powers of $\alpha_h$ coming from the efficiency factors $\kappa$ given in Eq.(\ref{eq: eff_factor_SW}). As previously seen in Eq.(\ref{eq: kappa_3_1}), we have $\kappa_3\approx \kappa_1$, which leads to a further simplification of the above expression. We find that Eq.(\ref{eq:omega31ratio}) becomes
\[
\frac{\Omega_{{\rm GW},3,obs}^{(p)}}{\Omega_{{\rm GW},1,obs}^{(p)}} \approx (1+\delta)^6\,\left(1-\frac{8\,\log(1+\delta)}{(S_3/T)_1}\,g(\kappa_1,\,\eta)\right)
\label{eq: omega1vs3}
\]
where 
\[
g(\kappa_1,\eta)\approx\left(2\frac{(\log(f_\beta(\kappa_1)))'}{(\log(f_S(\kappa_1)))'}-\left(\frac{3}{2}+\rm{eff}\right)\,\frac{(\log(f_\alpha(\kappa_1)))'}{(\log(f_S(\kappa_1)))'}+\frac{\frac{1}{2}-\rm{eff}}{(\log(f_S(\kappa_1)))'\,(\eta-\kappa_1)}\right)
\]
and 
\[\frac{8\,\log(1+\delta)}{(S_3/T)_1}\,g(\kappa_1,\,\eta) \ll 1
\] 
Hence, we see that the dominant contribution comes from the $(1+\delta)^6$ factor. Therefore, we can get a relative enhancement of roughly two orders of magnitude in the peak GW amplitudes between phase 1 and phase 3. 

In principle, a similar analysis could be conducted on the GW peak amplitudes for phase 2 and phase 3. In the case of maximal separation, we find that the amplitude ratio depends proportionally to the model parameter $\eta$ (in addition to the $\delta$ and $\xi_{0/f}$ values). However, the explicit functional forms of the amplitudes are not straightforward to express due to the different dominant GW production mechanisms involved in phase 2 and phase 3.

\section{Results}
Following the discussions in the previous sections we arrive at our GW spectrum plots. 
The three peaks in the overall GW spectrum correspond to three different phase transitions occurring at decreasing temperatures in the early universe. 
The size and shape of the GW spectrum is dependent on various phase transition parameters as well as the cosmological parameters viz.,$\{\xi_{0/f},\,\delta,\,T_0\}$, as qualitatively described in the previous section. 

We compare
our predictions to the power-law integrated sensitivity (PLIS) curves corresponding to a
signal-to-noise ratio (SNR) threshold of 1 for several future GW experiments: the spaced-based interferometers LISA \cite{Caprini:2019egz}, BBO \cite{Crowder:2005nr}, Ultimate-DECIGO
\cite{Braglia:2021fxn}, and µAres \cite{Sesana:2019vho}, and also include estimates for astrophysical foregrounds coming from galactic \cite{Robson:2018ifk} and extragalactic
compact binaries (CB) \cite{ Farmer:2003pa}.

\begin{figure}[H]
\centering

\includegraphics[width=\linewidth]{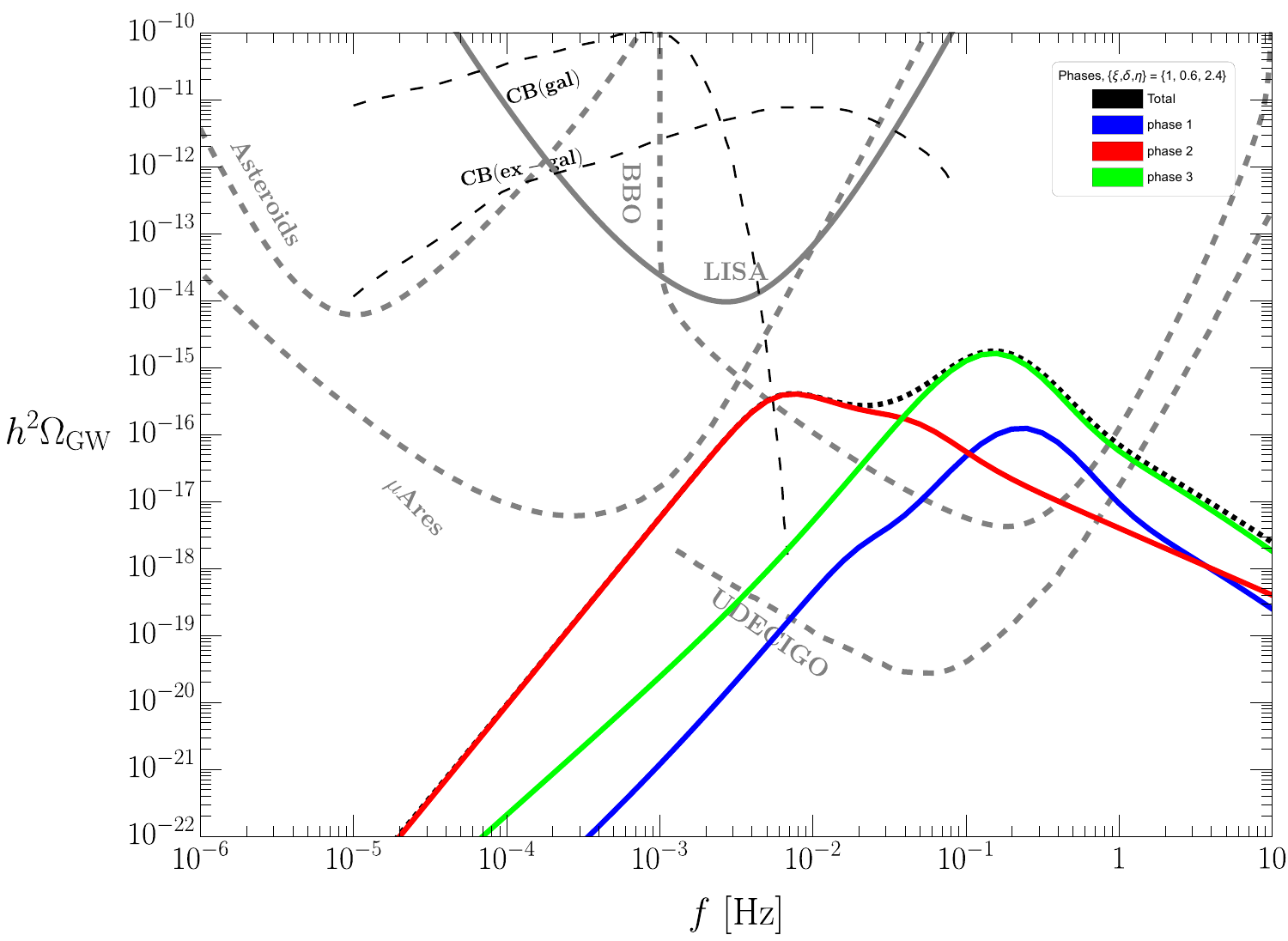} 

 \caption{ The GW spectrum corresponding to the three phases : Phase 1 (blue), Phase 2 (red) and Phase 3 (green) for a thermal kick of $\delta = 0.6$, $\xi = 1$, hidden sector scale $T_0 = 450 \,\rm{GeV}$ model dependence $\eta = 2.4$. Also
shown are the PLIS curves for upcoming experiments LISA (solid gray) and proposed experiments µAres, asteroid laser ranging, and
BBO (dashed gray). The PLIS curves for LISA, and BBO are adopted from \cite{Schmitz:2020syl, Batell:2023wdb} but
scaled to observation times of 3 yrs for LISA \cite{Caprini:2019egz}, and 4 yrs for BBO \cite{Crowder:2005nr}.
The µAres PLIS is taken from \cite{Sesana:2019vho}, scaled to SNR = 1. For the asteroid ranging proposal, we
adopt the strain sensitivity given in \cite{Fedderke:2021kuy} and calculate the PLIS curve using the procedure outlined
in \cite{Caprini:2019egz} for SNR = 1 and assumed an experiment duration of 7 yrs. For UltimateDECIGO (UDECIGO) we have adopted the PLIS in \cite{Braglia:2021fxn}. Black dashed lines represent foregrounds
from galactic and extragalactic compact binaries (CB)\cite{Robson:2018ifk, Farmer:2003pa}. Table \ref{tab:GW_para_xi_1} contains the information about the GW parameters corresponding the plots.}
\label{fig:xi1}
\end{figure}

Figure~\ref{fig:xi1} illustrates the case with $\xi_0=1$, $\delta=0.6$, $\eta=2.4$, and $T_0=450~\rm{GeV}$. We focus on the scenario where phase 2 and phase 3 peaks are maximally separated, occurring at $\kappa_1 \approx \kappa_3 = 1.55$, $\kappa_2 = 2.12$, and $\kappa_i = 2.17$, yielding $\tilde{\kappa}(\kappa_2,\kappa_i) = 1.61$. The phase transition parameters are listed in Table \ref{tab:GW_para_xi_1}, where $T_N^{SM}$ denotes the Standard Model (SM) temperature at each transition, which correlates with the Hubble value due to the SM-dominated energy density. The frequency peak of phase 2 is about an order of magnitude lower than that of phase 3, representing the maximum separation achievable for polynomial potentials.

Table \ref{tab:GW_para_xi075} presents phase transition parameters for $\xi_0=0.75$, $\eta=2.6$, and $T_0=350~\rm{GeV}$, comparing two scenarios: $\delta=0.57$ and $\delta=0.93$. Lowering both $T_0$ and $\xi_0$ in this scenario isolates impact of $\xi$ through the $\alpha$ parameter, keeping  the redshift effects to be similar.
A reduced $\xi_0$ leads to lower $h^2\,\Omega_P$ (peak amplitude) values for phase 1 in Table \ref{tab:GW_para_xi_1} relative to Table \ref{tab:GW_para_xi075} (see Eq.~(\ref{eq: alpha_xi})). Additionally, the influence of model parameters (via $\eta$) appears less significant than cosmological effects, as confirmed in both tables.

Both tables support the discussion on the impact of varying energy injection values ($\delta$): larger $\delta$ results in stronger GW amplitudes in phases 2 and 3 compared to phase 1, primarily due to higher $\xi_f$ and, therefore, larger $\alpha$ values via Eq.(\ref{eq: alpha_xi}). The tables also show that the peak frequencies for phases 1 and 3 are similar, with their ratio differing slightly from unity by a factor proportional to $\log(1+\delta)$, consistent with our analytical formula in Eq.(\ref{eq: fp3_fp1}). Additionally, both tables confirm the scaling prediction from Eq.(\ref{eq: omega1vs3}), where the predominant $h^2\,\Omega_P$ scaling is due to the factor $(1+\delta)^6$ for phase 3 vs phase 1. 

Recent literature \cite{guo2024precise} has raised questions about the standard semi-analytical estimates for sound wave-sourced GWs. Incorporating the new parameterization of~\cite{guo2024precise}, we found that the peaks for phases 1 and 3 shift to lower frequencies and smaller amplitudes, with the spectrum broadening due to two distinct peak frequencies. Despite these adjustments, the overall three-peak structure remains.

We would like to point out that we have focused on the scenario where the rate of energy injection is faster than the Hubble rate. In cases where the energy injection is adiabatic, the field will track its temperature-dependent minimum. The three transitions will then be $0\rightarrow \phi_c$ (Phase 1), $\phi_c\rightarrow 0$ (Phase 2), $0\rightarrow \phi_c$ (Phase 3). Thus, the first and third transitions are similar to those in the fast injection scenario, but the second is different. The key distinction between the three GW peaks would then be the redshift at which each transition occurs, assuming similar wall velocities.

As previously discussed, for larger energy injections, it is possible that the GW spectrum from phase 1 may be entirely submerged beneath that of phase 3, rendering it unobservable. In such a case, the GW spectra from phases 2 and 3 would still be detectable, offering valuable insights. This outcome would lead to an interpretation similar to the reheating scenario outlined in \cite{Buen-Abad:2023hex}, where the field begins from a non-zero vacuum at $T > T_c$, undergoes an inverse transition during heating in the hidden sector, and then transitions normally as the hidden sector cools. Both transitions would reflect the hidden sector temperature relative to the Standard Model, characterized by $\xi_f$, though information about $\delta$ would no longer be retained.

\begin{table}[!h]
\begin{center}
\renewcommand{\arraystretch}{1.5}
\begin{tabularx}{\textwidth}{|>{\centering\arraybackslash}X|>{\centering\arraybackslash}X|>{\centering\arraybackslash}X|>{\centering\arraybackslash}X|>{\centering\arraybackslash}X|>{\centering\arraybackslash}X|>{\centering\arraybackslash}X|>{\centering\arraybackslash}X|}
\hline
$\xi=1$ & $\alpha$ & $\beta/H$ & $\alpha_h$ & $T_N^{\text{SM}}$ (GeV) & $f_P$ (Hz) & $h^2\,\Omega_P$ & $\kappa$ \\
\hline
I & 0.004 & 885.7 & 0.074 & 751.3 & 0.23 & $1.2 \times 10^{-16}$ & 1.545 \\
\hline
II($\delta=0.6$) & 0.003 & 256.9 & 0.013 & 832.7 & 0.008 & $3.5 \times 10^{-16}$ & 1.61 \\
\hline
II($\delta=0.9$) & 0.005 & 257.8 & 0.013 & 696.7 & 0.007 & $9.8 \times 10^{-16}$ & 1.615 \\
\hline
III ($\delta =0.6$) & 0.018 & 909.9 & 0.071 & 477.3 & 0.15 & $1.5 \times 10^{-15}$ & 1.55 \\
\hline 
III ($\delta=0.9$) & 0.03 & 917.3 & 0.07 & 400.75 & 0.137 & $3.3 \times 10^{-15}$ & 1.55 \\ \hline
\end{tabularx}
\end{center}
\caption{Phase transition parameters for the different phases for $\xi = 1,\, \eta = 2.4,\,T_0=450 \rm{GeV}$ and for two different choices of $\delta = 0.6$ and $\delta=0.9$ .}
\label{tab:GW_para_xi_1}
\end{table}

\begin{table}[!h]
\begin{center}
\renewcommand{\arraystretch}{1.5}
\begin{tabularx}{\textwidth}{|>{\centering\arraybackslash}X|>{\centering\arraybackslash}X|>{\centering\arraybackslash}X|>{\centering\arraybackslash}X|>{\centering\arraybackslash}X|>{\centering\arraybackslash}X|>{\centering\arraybackslash}X|>{\centering\arraybackslash}X|}
\hline
$\xi = 0.75$ & $\alpha$ & $\beta/H$ & $\alpha_h$ & $T_N^{\text{SM}}$ (GeV) & $f_P$ (Hz) & $h^2\,\Omega_P$ & $\kappa$  \\
\hline
I & 0.0009 & 1070.5 & 0.055 & 941.46 & 0.145 & $1.1 \times 10^{-17}$ & 1.546 \\
\hline
II ($\delta = 0.57$) & 0.002 & 429.8 & 0.017 & 839.3 & 0.09 & $2.5 \times 10^{-17}$ & 1.61 \\
\hline
II ($\delta = 0.93$) & 0.003 & 435.62 & 0.017 & 565.1 & 0.009 & $1.2 \times 10^{-16}$ & 1.62 \\
\hline
III ($\delta = 0.57$) & 0.006 & 1108.8 & 0.055 & 565.02 & 0.1 & $1.4 \times 10^{-16}$ & 1.55 \\
\hline 
III ($\delta = 0.93$) & 0.01 & 1126.46 & 0.055 & 379.9 & 0.08 & $4.0 \times 10^{-16}$ & 1.55 \\
\hline
\end{tabularx}
\end{center}
\caption{Phase transition parameters for different phases at $\xi_0 = 0.75$ with two choices of $\delta = 0.57$ and $\delta = 0.93$ for $\eta = 2.6$ and $T_0 = 350 \,\mathrm{GeV}$.}
\label{tab:GW_para_xi075}
\end{table}
\section{Conclusion}

In this work, we have explored the potential of gravitational waves (GWs) to probe the pre-BBN era of the early universe through multiple phase transitions as a result of energy injection. Our analysis of the phase transition parameters highlights the effects of such early-universe energy injection, with the gravitational wave (GW) spectrum dependent on three key cosmological parameters: $\xi$, the temperature ratio between hidden and SM sectors, where $\xi$ is extracted by analyzing the difference between the peak frequencies of phase 2 and phase 3 GWs, $\delta$, quantifying the energy injection and is determined by comparing the phase 3 and phase 1 transitions, and $T_0$, the scale of hidden sector which can be  inferred from the peak frequency location. 

We demonstrate that energy injections in the hidden sector can lead to three distinct phase transitions, resulting in a unique three-peak GW spectrum. Of the three phase transitions, the first and third are qualitatively similar, differing mainly in redshift, while the second is fundamentally distinct, occurring in the opposite direction with a different broken phase. This net spectrum is broader and more complex than a typical single-peak spectrum, making it distinguishable with a combination of upcoming and planned GW observatories.
The two additional peaks, beyond the standard one get more enhanced with increasing energy injection into the hidden sector. We use a polynomial-like potential to explicitly show how the framework works, but the three peak structure is not limited to those class of models and can be generalized to other kinds of potentials as well.

Notably, the distinctive feature of multiple peaks in the GW spectrum appears regardless of the value of $\xi_{0/f}$, the mass scale of the hidden sector, or the specific class of models considered. This suggests that hidden sectors generating gravitational waves provide a robust tool for investigating a wide range of new physics scenarios in the pre-BBN era.

Moreover, the ability to detect and analyze these GW signals opens up exciting possibilities for understanding the dynamics of the early universe beyond the reach of traditional cosmological probes. By identifying the unique signatures of different phase transitions, future GW observations could offer unprecedented insights into the hidden sector's properties and the fundamental physics governing the early universe.

\section*{Acknowledgments} 

We thank Joachim Kopp, Arnab Dasgupta, Barmak Shams Es Haghi, Simone Blasi, Jae Hyeok Chang, Akshay Ghalsasi and Rudin Petrossian-Byrne for helpful discussions and correspondence. The work of BD and MR is supported by the U.S. Department of Energy Grant DE-SC0010813.  JBD acknowledges support from the National Science Foundation under grant
no. PHY-2412995.

 \newpage
 \appendix
\section{Reheating scenarios}
\subsection{Moduli decay to hidden sector}\label{app: moduli}

We can do a rough estimation for the case of a moduli field $\chi$ coupled to the hidden sector \footnote{We can take a more general case where it is coupled to both sectors, but still the relative change in temperature would be more for the hidden sector as compared to SM.}. As the decay rate $\Gamma_\chi$ \cite{Dutta:2009uf} of the moduli field becomes comparable to the Hubble, the field will decay to hidden sector and provide an energy boost.
For the decay to happen at the hidden sector temperature of roughly $100$ GeV \footnote{The scale of hidden sector which is sensitive to the energy injection is dictated by the moduli mass which gives the scale when it decays. }, we have
\begin{align}
    & \Gamma_\chi \approx \frac{m_\chi^3}{M_{pl}^2} \approx H = \frac{\sqrt{\frac{1}{3} \rho_R (T)}}{M_{pl}} \\ \nonumber
    \implies & m_\chi \approx 2.4\times 10^8\,\rm GeV \times \,\left(\frac{T}{100\,\rm GeV} \times \frac{0.1}{\xi}\right)^{2/3} 
\end{align}
The energy density  stored in the field $\chi$ near its oscillation is roughly given in terms of its initial field amplitude. Demanding that it is subdominant to the SM we get,
\begin{align}
    & \rho_\chi \approx m_\chi^2\,\chi_i^2 < \rho_{SM} \\ \nonumber
    \implies & \chi_i < 2.47 \times 10^{-2} \,\rm GeV\times \left(\frac{T}{100 \,\rm GeV}\times \frac{0.1}{\xi}\right)^{4/3} 
\end{align}
The amount of energy injection to PBH is then given by,
\begin{align}
    & \rho_h' = \rho_h + \rho_\chi \\ \nonumber 
    \implies & T_h' = T_h\left(1+\frac{30\,m_\chi^2\chi_i^2}{\pi^2\,g_h^*\,T_h^4}\right)^{1/4}
    \\ \nonumber
    \implies  
   & \delta = \left(1+\frac{30\,m_\chi^2\chi_i^2}{\pi^2\,g_h^*\,T_h^4}\right)^{1/4}-1 
\end{align}
Small $\delta < 1$ corresponds to smaller moduli initial field value. For example, for $\delta \approx 0.4$, we can have $\chi_i \approx 4\times 10^{-5} \,\rm GeV$,
\[
\delta \approx \frac{3}{8} \times \left(\frac{2.4\times 10^8 \times \chi_i }{10^4 \,\rm GeV}\right)^2 \approx 0.4
\]
Larger values of $\delta > 1$, which still obey the SM being the dominant contribution to the energy density, correspond to slightly larger initial field value; e.g., $\delta \approx 3$ corresponds to $\chi_i \approx 4.63 \times 10^{-4} \,\rm GeV$ as seen in
\[
\delta \approx 1.2 \times \frac{\sqrt{2.4\times 10^8 \,\rm GeV\times \chi_i} }{100\,\rm GeV} - 1 \approx 3
\]

Since we consider the case where the moduli field is only coupled to the hidden  sector, it implies that, post-decay, the ratio of SM and hidden sector temperatures is given by,
\[
\xi_f = \xi(1+\delta)
\]
\subsection{PBH reheating}\label{app: PBH}
Another instance for energy dumping in the early universe happens via PBH reheating. We emphasize that the impact of PBH reheating will be more on the hidden sector than the SM due to smaller energy content. 
PBH evaporation leads to entropy dilution, which can be estimated by following arguments in \cite{ Bernal:2021yyb} based on
conservation of energy before and after PBH evaporation. We see that
\begin{align}
   & \rho_R(T_f) + M_{PBH}\,n_0\frac{s(T_{f})}{s(T_{0})} = \rho_R(T_f') \\ 
    \implies & \rho_{SM}(T_{SM}) + \eta\,s(T_{SM})\frac{\rho_R(T_0)}{s(T_{0})} = \rho_{SM}(T_{SM}'), \\ \nonumber
    & \rho_h(T_h) + \eta\,s(T_{h})\frac{\rho_R(T_0)}{s(T_{0})} = \rho_h(T_H') 
\end{align}
where we used $\eta \rho_R(T_0) = M_{PBH}\,n_0,\,\rho_R(T_f) = \rho_h(T_h) + \rho_{SM}(T_{SM}),\,s(T_f)=s(T_{SM})+s(T_h)$, and we can simplify it further by using $\rho_R(T_0)/s(T_{0}) = 3\,T_0/4$, 
\begin{align}
     T_{SM}' &= T_{SM}\left(1+\frac{\eta\,T_0}{T_{SM}}\right)^{1/4} \\ 
     T_{h}' &= T_{h}\left(1+\frac{\eta\,T_0}{T_h}\right)^{1/4}
\end{align}
It is clear from the above equations that the change to the SM temperature is $\xi$ suppressed relative to the hidden sector. 
Following \cite{Bernal:2022pue}, PBHs are formed roughly at the SM temperature given as \footnote{ A more precise calculation will involve defining an effective temperature $\tilde{T}_0 = (g_h^* \xi^4 + g_{SM}^*)^{1/4}T_0 $, and since $\xi <1 $ and $g_h^*<g_{SM}^*$, we have $\tilde{T_0}\approx T_0$, where $T_0$ corresponds to the tempeature of the SM at the time of PBH formation.},
\[
T_0^{SM} \approx 1.85 \times 10^{13} {\rm GeV} \left(\frac{5.3 \times 10^4~{\rm g} }{M_{BH0}}\right)^{1/2}
\]
and the evaporation temperature is given as,
\[
T_{ev}^{SM} \approx  10^3 {\rm GeV} \times \left(\frac{5.3 \times 10^4~{\rm g} }{M_{BH0}}\right)^{3/2}
\]
The SM temperature of $\rm TeV$ corresponds to $T_h \approx 100 \rm GeV$ for $\xi = 0.1$. For small $\delta < 1$, 
\[
\delta \approx 0.45\times \frac{\eta}{ 10^{-11}}\times \frac{0.1}{\xi} \times \left(\frac{M_{BH0}}{5.3\times 10^4 ~{\rm g}}\right)
\]
For larger delta values, $\delta > 1$, we need larger $\eta$ values\footnote{Impact on SM temperature for $\delta < 4$ is small since $\delta_{SM}\approx \delta \times \xi / 4<0.2$ for $\xi < 0.2$, and thus can be ignored.}. For instance, for $\delta \approx 3$ we have,
\[
\delta \approx 4\left(\frac{\eta}{ 1.39 \times 10^{-9}}\times \frac{0.1}{\xi} \times \left(\frac{M_{BH0}}{5.3\times 10^4 g}\right)\right)^{1/4} -1 
\]

For the SM, we can get the analogous change in temperature, which is
\[
T_{SM}'=\frac{T_{h}}{\xi}\left(1+\xi\,((1+\delta)^4-1)\right)^{1/4}
\]
This implies that the new ratio between the hidden sector and the SM after the injection is given as,
\[
\xi_f = \frac{\xi(1+\delta)}{\sqrt[4]{\left((\delta +1)^4-1\right) \xi +1}} \approx \xi(1+\delta),\quad \xi\ll 1 
\]

 \section{Phase transition parameters}
 \label{app:parameters}

\subsection{Phase transition milestones}
We can briefly mention the key values for the field. Initially at high temperature, $T>T_1$, there is only one minimum at $\phi=0$. As the  universe cools, for $T=T_1$, one finds a second minimum provided by,
\[
T_1^2= \frac{T_0^2}{1-\frac{9E^2}{8\lambda\,D}},\quad \phi_1 = \frac{3E\,T_1}{2\lambda}
\label{eq : T1}
\]
The two minima become degenerate at the critical temperature, $V(0,T_c)=V(\phi_c,T_c)$, given by,
\[
T_c^2= \frac{T_0^2}{1-\frac{E^2}{\lambda\,D}},\quad \phi_c = \frac{2E\,T_c}{\lambda}
\]
As the universe cools down further, at $T=T_0$, $\phi=0$ ceases to be a minimum, and the only remaining minimum is given by, 
\[
\phi_0 = \frac{3E\,T_0}{\lambda}
\]

 \subsection{Euclidean Action}
 The tunnelling action is an important quantity required in order to calculate both the rate of transition as well as the nucleation temperature.
For simple polynomial-like potentials as seen in Eq.(\ref{eq:V}), there is an accurate semi-analytical approximation of the tunneling
action \cite{Adams:1993zs, Ellis:2020awk},
\[
\frac{S_3}{T} = \frac{2\sqrt{2}\,E}{\lambda^{3/2}} \,\frac{8\,\pi\,\kappa^{3/2}(8.2938-5.533\kappa+0.818\kappa^2)}{81(2-\kappa^2)}
\]
where 
\[
\kappa = \frac{2\lambda\,D\left(T^2-T_0^2\right)}{E^2\,T^2}, \quad 0\leq\kappa\leq2
\label{eq: kappa}
\]

The results are applicable when the false vacua is defined around $\phi = 0$. For phase 2, we can redefine the fields such that $\phi_i$ becomes zero-valued. The field transformation needed is $\tilde{\phi}=\phi-\phi_i$. The potential is then given by,
\begin{align}
    \tilde{V}(\tilde{\phi}) & =  \tilde{D}(T^2-T_0^2)\tilde{\phi}^2-\tilde{E}\,T\,\tilde{\phi}^3+\frac{\lambda}{4}\tilde{\phi}^4 \\
 & = \tilde{D}(T^2-T_0^2)(\phi-\phi_i)^2-\tilde{E}\,T\,(\phi-\phi_i)^3+\frac{\lambda}{4}(\phi-\phi_i)^4 
\end{align}
This should be matched to our original potential, since $V(\phi) = \tilde{V}(\tilde{\phi})$. The reasoning behind this is that we can start from our original potential and add and subtract terms to get the new function which is centered around $\tilde{\phi}$. Dropping the linear terms and the constant, we get, 
\begin{align}
   & \tilde{D} = D+ 
    \frac{3\phi_i(-2\,E\,T+\lambda \phi_i)}{2((T_i(1+\delta))^2-T^2_0)} = D\,\left(\frac{9-4\kappa(\tilde{T}_i)+3\,\sqrt{9-4\,\kappa(\tilde{T}_i)}}{2\,\kappa(\tilde{T}_i)}\right)  \label{eq: Phase_II_D}\\ & \tilde{E} = E -\frac{\lambda\phi_i}{T_i(1+\delta)} = E\left(\frac{1+\sqrt{9-4\,\kappa(\tilde{T}_i)}}{2}\right) \label{eq: Phase_II_E} \\ 
   & \tilde{\kappa} = \frac{2\,\lambda\,\tilde{D}}{\tilde{E}^2 }\left(1-\frac{T_0^2}{T^2}\right) = \kappa(T)\,\left( \frac{9-4\,\kappa(\tilde{T}_i)+3\sqrt{9-4\,\kappa(\tilde{T}_i)}}{\kappa(\tilde{T}_i)(5-2\,\kappa(\tilde{T}_i)+\sqrt{9-4\,\kappa(\tilde{T}_i)})} \right)
   \label{eq: Phase_II_kappa}
\end{align}
where $\tilde{T}_i=T_i(\delta+1)$ and $2<\kappa(\tilde{T}_i)<9/4$ for $T_c<\tilde{T}_i<T_1$. 
Also, we can redefine $\tilde{\phi}\rightarrow -\tilde{\phi}$ to reorient the second minimum corresponding to $\phi=0$ in terms of our older variables corresponding to the symmetric phase along the positive field values. This is equivalent to $\tilde{E}\rightarrow -\tilde{E}$. Thus, in this way we can use the standard results to calculate the phase transition parameters.

 \subsection{Nucleation Temperature}
  
  The nucleation temperature can be obtained via analytical estimates and is given by \cite{Athron:2023xlk}, 
\[
\Gamma/ H^4 \approx 1
\]
where,
\[
\Gamma = T^4\left(\frac{S_3}{2\pi\,T}\right)^{3/2}\exp^{-\frac{S_3}{2\pi\,T}}
\]
and $S_3$ is the Euclidean action.
\subsection{Strength of Phase Transition}

The strength of the phase transition is given by $\alpha$, defined as the ratio of the latent heat released in the transition to the critical energy density of the universe at the nucleation temperature. It is given as \cite{Athron:2023xlk},
\[
\alpha = \frac{\Delta (V-\frac{T}{4}\partial_T V)}{\rho_R}\biggr\rvert_{T=T_N}
\]
where $T_N$ is the nucleation temperature.
Starting from Eq.(\ref{eq:V}), we get,
\begin{equation}
        \alpha \approx  \frac{\Delta(2\,D\,(T_N^2-2\,T_0^2)\phi^2-3\,E\,T_N\,\phi^3+\lambda\,\phi^4)}{4\rho_R}
\end{equation}
 \subsection{Duration of phase transition}
 
 The rate of the phase transition, $\beta/H_0$, is given in terms of the bounce action. 
Following the definition of $\beta$, we get \cite{Athron:2023xlk},
\[
  \frac{\beta}{H_N}  = T\frac{d(S_3/T)}{dT}|_{T_N} 
  \label{eq:beta}
\]
where $T_0<T_N<T_c$ is the nucleation temperature.
 \subsection{Wall velocity}
 
The bubble wall velocity can be calculated by solving Boltzmann equations numerically \cite{Laurent:2022jrs}. We can classify the phase transitions into two categories:
1) deflagrations where $v_w \approx c_s$  or 2) ultrarelativistic detonations where $v_w \approx 1$. Physically, deflagrations can be achieved for weaker phase transitions, since the bubble walls are unable to overcome the friction generated by the ambient plasma and hence attain a terminal velocity. For stronger transitions, the bubble walls can enter the run-away phase, where the bubble walls reach ultra-relativistic speeds. There is ongoing debate in the literature regarding whether the bubble walls actually reach $v_w\approx 1$, but for our purposes we assume it will.  
Following \cite{Basler:2024aaf} the wall velocity can be well approximated as,
\begin{equation}
v_w = \begin{cases}
\sqrt{\frac{\Delta V}{\alpha \rho_R}} &\sqrt{\frac{\Delta V}{\alpha \rho_R}} < v_J\\
1 &\sqrt{\frac{\Delta V}{\alpha \rho_R}} > v_J
\end{cases}
\label{eq : wall_velocity}
\end{equation}
where the Jouguet velocity $v_J$ is given by,
\[
v_J = \frac{1}{1+\alpha}\left(\sqrt{\frac{1}{3}}+\sqrt{\alpha\left(\frac{2}{3}+\alpha\right)}\right)
\]
In the above expressions, all quantities are  calculated at $T_N$.

\section{Gravitational wave spectrum}

\label{app:GW}

The production mechanisms can be demarcated into two separate regimes, namely runaway transitions $\alpha >\alpha_\infty$ and non-runaway transitions $\alpha<\alpha_\infty$, where \cite{Ellis:2019oqb}
\[
\alpha_\infty = \frac{\Delta p_{LO}}{\rho_R} = \sum_i \frac{c_i\, \Delta m_i^2}{24\,\rho_R}
\]
where $c_i = 1 $ for bosons and $1/2$ for fermions. $\alpha_\infty$ corresponds to the friction pressure due to mass gain as particles enter the true vacuum. There are higher order corrections as well, which give rise to a upper bound on $\gamma_w$ for the wall velocity \cite{Azatov:2023xem}.

\subsection{Runaway transition (Phase 2)}
A semi-analytical form for the gravitational wave power spectrum at emission for the case of runaway transitions is given by~\cite{Huber:2008hg,Caprini:2009yp,Hindmarsh:2015qta}, 
\[
\Omega_{GW}(f)\simeq
 \sum_i \mathcal N_i \, \Delta_i(v_w) \,\left(\frac{\kappa_i \, \alpha}{1+\alpha}\right)^{p_i}
    \left(\frac{H}{\beta}\right)^{q_i}  s_i(f/f_{p,i}) \,,
    \label{eq:OmegaGW-emission}
\]
where $i\in \{$BW, SW, turb$\}$ runs over contributions from bubble walls (BW), sound waves (SW) and turbulent effects (turb).
We see from Eq.\eqref{eq:OmegaGW-emission} that the GW spectrum depends on the phase transition strength parameters $\alpha_{\rm tot}$, the phase transition duration parameter $\beta/H$, the bubble wall velocity $v_{\rm w}$, and the efficiency factors  $\kappa_{\rm BW}$ (fraction of the vacuum energy carried by the bubble walls during collision) and $\kappa_{\rm SW}$ (energy fraction transferred to plasma bulk motion). 
In the presence of negative friction due to mass loss of particles crossing the walls, the efficiency factor for bubble walls $\kappa_{BW}$ for phase 2 is extracted from a detailed analysis \cite{Barni:2024lkj}. A naive application of the efficiency factor in terms of $\alpha_\infty$ would yield $\kappa_{\rm BW} >1$ \cite{Buen-Abad:2023hex}, which is contrary to the second law of thermodynamics prohibiting over-efficient extraction of energy from the system. The sound wave efficiency factor is given by~\cite{Ellis:2019oqb},
\[
\kappa_{\rm SW} = (1-\kappa_{BW})\,\frac{\alpha_{eff}}{0.73+0.083\sqrt{\alpha_{eff}}+\alpha_{eff}},\quad \alpha_{eff} = \alpha_h(1-\kappa_{\rm BW})
\]
Thus, for our Phase 2 scenario, the bulk of the GW are being sourced from the bubble wall collisions.

For the other quantities appearing in Eq.\eqref{eq:OmegaGW-emission} we use the results from Refs.~\cite{Caprini:2019egz,Huber:2008hg}. For the normalization factors, we have $(N_{\rm BW},N_{\rm SW},N_{\rm turb}) = (1, 0.159, 20.1)$.
The velocity factor takes into account a potential suppression due to the wall velocity.  
The exponents are given by $(p_{\rm BW},p_{\rm SW},p_{\rm turb}) = (2,2,3/2)$ and $(q_{\rm BW},q_{\rm SW}, q_{\rm turb}) = (2,1,1)$. The velocity factors, spectral shape functions, and corresponding peak frequencies are taken to be \cite{Breitbach:2018ddu}
\begin{align}
    \label{eq:spectrum}
    & \Delta_{\rm BW}=\frac{0.11 v^{3}_{\rm w}}{(0.42 + v^{2}_{\rm w})},\quad f_{\rm p, BW} = \frac{0.62\beta}{1.8-0.1v_{\rm w}+v_{\rm w}^2},\quad s_{\rm BW}(x) = \frac{3.8 \, x^{2.8}}{1+2.8 \, x^{3.8}},   \nonumber \\ & \Delta_{\rm SW}=v_{\rm w}\min\,(1,H_*\tau_\mathrm{sh}),\qquad  f_{\rm p, SW} = \frac{2\beta}{\sqrt{3}v_{\rm w}},\qquad s_{\rm SW}(x) =x^{3}\left(\frac{7}{4+3 \, x^{2}}\right)^{7/2},
    \nonumber \\ & \Delta_{\rm turb}=v_{\rm w},\qquad  f_{\rm p, turb} = \frac{3.5\beta}{2v_{\rm w}},\qquad s_{\rm turb}(x) =\frac{x^{3}}{(1+x)^{11/3}(1+8\pi\,x \frac{f_{\rm p, turb}}{H})}.
\end{align}
For the sound wave contribution with $H_*\tau_\mathrm{sh} \neq 1$\footnote{$H_*\tau_\mathrm{sh}<1$ corresponds to a short lasting transition and $H_*\tau_\mathrm{sh}>1$ for a long lasting transition.}, this additional suppression factor \cite{Ellis:2020awk,Guo:2020grp} is given by, 
\[H_*\tau_\mathrm{sh} \approx \frac{3.38 \, {\rm max}[v_{\rm w}, c_{\rm s}]}{\beta / H} \sqrt{\frac{1+\alpha}{\kappa_{\rm  SW}\alpha}}\]
where $c_{\rm s} \approx 1/\sqrt{3}$ is the speed of sound in the relativistic plasma.

The turbulence efficiency factor is given by 
\begin{align}
    \kappa_\mathrm{Turb} = \epsilon\kappa_\mathrm{SW}\,,
\end{align}
where $\epsilon $ is given by \cite{Hindmarsh:2017gnf},
\begin{align}
    \epsilon=\left(1-\text{min}\left(H_*\tau_\mathrm{sh}, 1\right)\right)^{2/3}
\end{align}

\subsection{Non-runaway transition (Phase 1 and Phase 3)}
For non-runaway transitions, the contribution from bubble collisions is small and there exist a lot of uncertainties, thus we will not consider it \cite{Basler:2024aaf}.
The GW power spectrum originating from sound waves and
turbulence is still given by the spectral functions defined in the previous section. The difference compared to the runaway case comes via the sound wave efficiency factor given by \cite{Espinosa:2010hh,Basler:2024aaf}
\begin{align}
 	\kappa_\mathrm{SW}&=
	 \begin{cases}
      \frac{c_s^{11 / 5} \kappa_A \kappa_B}{\left(c_s^{11 / 5}-v_w^{11 / 5}\right) \kappa_B+v_w c_s^{6 / 5} \kappa_A}, & \text{if}\ v_w<c_s \\
      \kappa_B+\left(v_w-c_s\right) \delta \kappa+\frac{\left(v_w-c_s\right)^3}{\left(v_J-c_s\right)^3}\left[\kappa_C-\kappa_B-\left(v_J-c_s\right) \delta \kappa\right], & \text{if}\ c_s<v_w<v_J \\
      \frac{\left(v_J-1\right)^3 v_J^{5 / 2} v_w^{-5 / 2} \kappa_C \kappa_D}{\left[\left(v_J-1\right)^3-\left(v_w-1\right)^3\right] v_J^{5 / 2} \kappa_C+\left(v_w-1\right)^3 \kappa_D}, & \text{if}\ v_J<v_w 
	\end{cases}
 \label{eq: eff_factor_SW}
\end{align}
with
\begin{align}
	\kappa_A &\simeq v_w^{6 / 5} \frac{6.9 \,\alpha_h}{1.36-0.037
                   \sqrt{\alpha_h}+\alpha_h}, &  \kappa_B \simeq
  &\frac{\alpha_h^{2 / 5}}{0.017+\left(0.997+\alpha_h\right)^{2 / 5}},
    \nonumber \\
	\kappa_C &\simeq
                   \frac{\sqrt{\alpha_h}}{0.135+\sqrt{0.98+\alpha_h}}, &
                                                                     \kappa_D
  &\simeq \frac{\alpha_h}{0.73+0.083 \sqrt{\alpha_h}+\alpha_h}, \nonumber\\
	\delta \kappa &\simeq -0.9 \log \frac{\sqrt{\alpha_h}}{1+\sqrt{\alpha_h}}.
\end{align}

We can do a rough estimate to show that the production from sound waves is always more dominant than the one from turbulence for $\beta/H \in \mathcal{O}(10)-\mathcal{O}(100)$. At the peak frequency of each\footnote{ The spectral shape function for turbulence is maximized at $f \approx 1.2 f_P$, and is roughly given as \\$s_{turb}(1.2)\approx \frac{0.0018\,v_w}{\beta/H}$.}, the corresponding ratio can be given as,
\begin{align}
\frac{\Omega_{turb}}{\Omega_{SW}} & \approx \frac{20.1}{0.159}\times  \frac{\beta}{H} \times \frac{1}{3.38 \, {\rm max}[v_{\rm w}, c_{\rm s}]} \times s_{turb}(1.2) \times (1-H\,\tau_{sh}) \\ \nonumber
& \approx 37.4 \times  \frac{\beta}{H} \times  \frac{0.0018}{\beta/H} \times \frac{v_w}{{\rm max}[v_{\rm w}, c_{\rm s}]} \approx 0.079 \times (1-H\,\tau_{sh}) \ll 1
\end{align}


\bibliography{main}

\begin{thebibliography}{73}%
\makeatletter
\providecommand \@ifxundefined [1]{%
 \@ifx{#1\undefined}
}%
\providecommand \@ifnum [1]{%
 \ifnum #1\expandafter \@firstoftwo
 \else \expandafter \@secondoftwo
 \fi
}%
\providecommand \@ifx [1]{%
 \ifx #1\expandafter \@firstoftwo
 \else \expandafter \@secondoftwo
 \fi
}%
\providecommand \natexlab [1]{#1}%
\providecommand \enquote  [1]{``#1''}%
\providecommand \bibnamefont  [1]{#1}%
\providecommand \bibfnamefont [1]{#1}%
\providecommand \citenamefont [1]{#1}%
\providecommand \href@noop [0]{\@secondoftwo}%
\providecommand \href [0]{\begingroup \@sanitize@url \@href}%
\providecommand \@href[1]{\@@startlink{#1}\@@href}%
\providecommand \@@href[1]{\endgroup#1\@@endlink}%
\providecommand \@sanitize@url [0]{\catcode `\\12\catcode `\$12\catcode `\&12\catcode `\#12\catcode `\^12\catcode `\_12\catcode `\%12\relax}%
\providecommand \@@startlink[1]{}%
\providecommand \@@endlink[0]{}%
\providecommand \url  [0]{\begingroup\@sanitize@url \@url }%
\providecommand \@url [1]{\endgroup\@href {#1}{\urlprefix }}%
\providecommand \urlprefix  [0]{URL }%
\providecommand \Eprint [0]{\href }%
\providecommand \doibase [0]{https://doi.org/}%
\providecommand \selectlanguage [0]{\@gobble}%
\providecommand \bibinfo  [0]{\@secondoftwo}%
\providecommand \bibfield  [0]{\@secondoftwo}%
\providecommand \translation [1]{[#1]}%
\providecommand \BibitemOpen [0]{}%
\providecommand \bibitemStop [0]{}%
\providecommand \bibitemNoStop [0]{.\EOS\space}%
\providecommand \EOS [0]{\spacefactor3000\relax}%
\providecommand \BibitemShut  [1]{\csname bibitem#1\endcsname}%
\let\auto@bib@innerbib\@empty
\bibitem [{\citenamefont {Abbott}\ \emph {et~al.}(2023)\citenamefont {Abbott} \emph {et~al.}}]{KAGRA:2021vkt}%
  \BibitemOpen
  \bibfield  {author} {\bibinfo {author} {\bibfnamefont {R.}~\bibnamefont {Abbott}} \emph {et~al.} (\bibinfo {collaboration} {KAGRA, VIRGO, LIGO Scientific}),\ }\bibfield  {title} {\bibinfo {title} {{GWTC-3: Compact Binary Coalescences Observed by LIGO and Virgo during the Second Part of the Third Observing Run}},\ }\href {https://doi.org/10.1103/PhysRevX.13.041039} {\bibfield  {journal} {\bibinfo  {journal} {Phys. Rev. X}\ }\textbf {\bibinfo {volume} {13}},\ \bibinfo {pages} {041039} (\bibinfo {year} {2023})},\ \Eprint {https://arxiv.org/abs/2111.03606} {arXiv:2111.03606 [gr-qc]} \BibitemShut {NoStop}%
\bibitem [{\citenamefont {Agazie}\ \emph {et~al.}(2023)\citenamefont {Agazie} \emph {et~al.}}]{NANOGrav:2023gor}%
  \BibitemOpen
  \bibfield  {author} {\bibinfo {author} {\bibfnamefont {G.}~\bibnamefont {Agazie}} \emph {et~al.} (\bibinfo {collaboration} {NANOGrav}),\ }\bibfield  {title} {\bibinfo {title} {{The NANOGrav 15 yr Data Set: Evidence for a Gravitational-wave Background}},\ }\href {https://doi.org/10.3847/2041-8213/acdac6} {\bibfield  {journal} {\bibinfo  {journal} {Astrophys. J. Lett.}\ }\textbf {\bibinfo {volume} {951}},\ \bibinfo {pages} {L8} (\bibinfo {year} {2023})},\ \Eprint {https://arxiv.org/abs/2306.16213} {arXiv:2306.16213 [astro-ph.HE]} \BibitemShut {NoStop}%
\bibitem [{\citenamefont {Antoniadis}\ \emph {et~al.}(2023)\citenamefont {Antoniadis} \emph {et~al.}}]{EPTA:2023fyk}%
  \BibitemOpen
  \bibfield  {author} {\bibinfo {author} {\bibfnamefont {J.}~\bibnamefont {Antoniadis}} \emph {et~al.} (\bibinfo {collaboration} {EPTA, InPTA:}),\ }\bibfield  {title} {\bibinfo {title} {{The second data release from the European Pulsar Timing Array - III. Search for gravitational wave signals}},\ }\href {https://doi.org/10.1051/0004-6361/202346844} {\bibfield  {journal} {\bibinfo  {journal} {Astron. Astrophys.}\ }\textbf {\bibinfo {volume} {678}},\ \bibinfo {pages} {A50} (\bibinfo {year} {2023})},\ \Eprint {https://arxiv.org/abs/2306.16214} {arXiv:2306.16214 [astro-ph.HE]} \BibitemShut {NoStop}%
\bibitem [{\citenamefont {Reardon}\ \emph {et~al.}(2023)\citenamefont {Reardon} \emph {et~al.}}]{Reardon:2023gzh}%
  \BibitemOpen
  \bibfield  {author} {\bibinfo {author} {\bibfnamefont {D.~J.}\ \bibnamefont {Reardon}} \emph {et~al.},\ }\bibfield  {title} {\bibinfo {title} {{Search for an Isotropic Gravitational-wave Background with the Parkes Pulsar Timing Array}},\ }\href {https://doi.org/10.3847/2041-8213/acdd02} {\bibfield  {journal} {\bibinfo  {journal} {Astrophys. J. Lett.}\ }\textbf {\bibinfo {volume} {951}},\ \bibinfo {pages} {L6} (\bibinfo {year} {2023})},\ \Eprint {https://arxiv.org/abs/2306.16215} {arXiv:2306.16215 [astro-ph.HE]} \BibitemShut {NoStop}%
\bibitem [{\citenamefont {Xu}\ \emph {et~al.}(2023)\citenamefont {Xu} \emph {et~al.}}]{Xu:2023wog}%
  \BibitemOpen
  \bibfield  {author} {\bibinfo {author} {\bibfnamefont {H.}~\bibnamefont {Xu}} \emph {et~al.},\ }\bibfield  {title} {\bibinfo {title} {{Searching for the Nano-Hertz Stochastic Gravitational Wave Background with the Chinese Pulsar Timing Array Data Release I}},\ }\href {https://doi.org/10.1088/1674-4527/acdfa5} {\bibfield  {journal} {\bibinfo  {journal} {Res. Astron. Astrophys.}\ }\textbf {\bibinfo {volume} {23}},\ \bibinfo {pages} {075024} (\bibinfo {year} {2023})},\ \Eprint {https://arxiv.org/abs/2306.16216} {arXiv:2306.16216 [astro-ph.HE]} \BibitemShut {NoStop}%
\bibitem [{\citenamefont {Hawking}(1971)}]{Hawking:1971ei}%
  \BibitemOpen
  \bibfield  {author} {\bibinfo {author} {\bibfnamefont {S.}~\bibnamefont {Hawking}},\ }\bibfield  {title} {\bibinfo {title} {{Gravitationally collapsed objects of very low mass}},\ }\href {https://doi.org/10.1093/mnras/152.1.75} {\bibfield  {journal} {\bibinfo  {journal} {Mon. Not. Roy. Astron. Soc.}\ }\textbf {\bibinfo {volume} {152}},\ \bibinfo {pages} {75} (\bibinfo {year} {1971})}\BibitemShut {NoStop}%
\bibitem [{\citenamefont {Carr}\ and\ \citenamefont {Hawking}(1974)}]{Carr:1974nx}%
  \BibitemOpen
  \bibfield  {author} {\bibinfo {author} {\bibfnamefont {B.~J.}\ \bibnamefont {Carr}}\ and\ \bibinfo {author} {\bibfnamefont {S.~W.}\ \bibnamefont {Hawking}},\ }\bibfield  {title} {\bibinfo {title} {{Black holes in the early Universe}},\ }\href {https://doi.org/10.1093/mnras/168.2.399} {\bibfield  {journal} {\bibinfo  {journal} {Mon. Not. Roy. Astron. Soc.}\ }\textbf {\bibinfo {volume} {168}},\ \bibinfo {pages} {399} (\bibinfo {year} {1974})}\BibitemShut {NoStop}%
\bibitem [{\citenamefont {Carr}(1975)}]{Carr:1975qj}%
  \BibitemOpen
  \bibfield  {author} {\bibinfo {author} {\bibfnamefont {B.~J.}\ \bibnamefont {Carr}},\ }\bibfield  {title} {\bibinfo {title} {{The Primordial black hole mass spectrum}},\ }\href {https://doi.org/10.1086/153853} {\bibfield  {journal} {\bibinfo  {journal} {Astrophys. J.}\ }\textbf {\bibinfo {volume} {201}},\ \bibinfo {pages} {1} (\bibinfo {year} {1975})}\BibitemShut {NoStop}%
\bibitem [{\citenamefont {Kawasaki}\ \emph {et~al.}(2000)\citenamefont {Kawasaki}, \citenamefont {Kohri},\ and\ \citenamefont {Sugiyama}}]{Kawasaki:2000en}%
  \BibitemOpen
  \bibfield  {author} {\bibinfo {author} {\bibfnamefont {M.}~\bibnamefont {Kawasaki}}, \bibinfo {author} {\bibfnamefont {K.}~\bibnamefont {Kohri}},\ and\ \bibinfo {author} {\bibfnamefont {N.}~\bibnamefont {Sugiyama}},\ }\bibfield  {title} {\bibinfo {title} {{MeV scale reheating temperature and thermalization of neutrino background}},\ }\href {https://doi.org/10.1103/PhysRevD.62.023506} {\bibfield  {journal} {\bibinfo  {journal} {Phys. Rev. D}\ }\textbf {\bibinfo {volume} {62}},\ \bibinfo {pages} {023506} (\bibinfo {year} {2000})},\ \Eprint {https://arxiv.org/abs/astro-ph/0002127} {arXiv:astro-ph/0002127} \BibitemShut {NoStop}%
\bibitem [{\citenamefont {Carr}\ \emph {et~al.}(2021)\citenamefont {Carr}, \citenamefont {Kohri}, \citenamefont {Sendouda},\ and\ \citenamefont {Yokoyama}}]{Carr:2020gox}%
  \BibitemOpen
  \bibfield  {author} {\bibinfo {author} {\bibfnamefont {B.}~\bibnamefont {Carr}}, \bibinfo {author} {\bibfnamefont {K.}~\bibnamefont {Kohri}}, \bibinfo {author} {\bibfnamefont {Y.}~\bibnamefont {Sendouda}},\ and\ \bibinfo {author} {\bibfnamefont {J.}~\bibnamefont {Yokoyama}},\ }\bibfield  {title} {\bibinfo {title} {{Constraints on primordial black holes}},\ }\href {https://doi.org/10.1088/1361-6633/ac1e31} {\bibfield  {journal} {\bibinfo  {journal} {Rept. Prog. Phys.}\ }\textbf {\bibinfo {volume} {84}},\ \bibinfo {pages} {116902} (\bibinfo {year} {2021})},\ \Eprint {https://arxiv.org/abs/2002.12778} {arXiv:2002.12778 [astro-ph.CO]} \BibitemShut {NoStop}%
\bibitem [{\citenamefont {Villanueva-Domingo}\ \emph {et~al.}(2021)\citenamefont {Villanueva-Domingo}, \citenamefont {Mena},\ and\ \citenamefont {Palomares-Ruiz}}]{Villanueva-Domingo:2021spv}%
  \BibitemOpen
  \bibfield  {author} {\bibinfo {author} {\bibfnamefont {P.}~\bibnamefont {Villanueva-Domingo}}, \bibinfo {author} {\bibfnamefont {O.}~\bibnamefont {Mena}},\ and\ \bibinfo {author} {\bibfnamefont {S.}~\bibnamefont {Palomares-Ruiz}},\ }\bibfield  {title} {\bibinfo {title} {{A brief review on primordial black holes as dark matter}},\ }\href {https://doi.org/10.3389/fspas.2021.681084} {\bibfield  {journal} {\bibinfo  {journal} {Front. Astron. Space Sci.}\ }\textbf {\bibinfo {volume} {8}},\ \bibinfo {pages} {87} (\bibinfo {year} {2021})},\ \Eprint {https://arxiv.org/abs/2103.12087} {arXiv:2103.12087 [astro-ph.CO]} \BibitemShut {NoStop}%
\bibitem [{\citenamefont {Moroi}\ and\ \citenamefont {Randall}(2000)}]{Moroi:1999zb}%
  \BibitemOpen
  \bibfield  {author} {\bibinfo {author} {\bibfnamefont {T.}~\bibnamefont {Moroi}}\ and\ \bibinfo {author} {\bibfnamefont {L.}~\bibnamefont {Randall}},\ }\bibfield  {title} {\bibinfo {title} {{Wino cold dark matter from anomaly mediated SUSY breaking}},\ }\href {https://doi.org/10.1016/S0550-3213(99)00748-8} {\bibfield  {journal} {\bibinfo  {journal} {Nucl. Phys. B}\ }\textbf {\bibinfo {volume} {570}},\ \bibinfo {pages} {455} (\bibinfo {year} {2000})},\ \Eprint {https://arxiv.org/abs/hep-ph/9906527} {arXiv:hep-ph/9906527} \BibitemShut {NoStop}%
\bibitem [{\citenamefont {Dutta}\ \emph {et~al.}(2009)\citenamefont {Dutta}, \citenamefont {Leblond},\ and\ \citenamefont {Sinha}}]{Dutta:2009uf}%
  \BibitemOpen
  \bibfield  {author} {\bibinfo {author} {\bibfnamefont {B.}~\bibnamefont {Dutta}}, \bibinfo {author} {\bibfnamefont {L.}~\bibnamefont {Leblond}},\ and\ \bibinfo {author} {\bibfnamefont {K.}~\bibnamefont {Sinha}},\ }\bibfield  {title} {\bibinfo {title} {{Mirage in the Sky: Non-thermal Dark Matter, Gravitino Problem, and Cosmic Ray Anomalies}},\ }\href {https://doi.org/10.1103/PhysRevD.80.035014} {\bibfield  {journal} {\bibinfo  {journal} {Phys. Rev. D}\ }\textbf {\bibinfo {volume} {80}},\ \bibinfo {pages} {035014} (\bibinfo {year} {2009})},\ \Eprint {https://arxiv.org/abs/0904.3773} {arXiv:0904.3773 [hep-ph]} \BibitemShut {NoStop}%
\bibitem [{\citenamefont {Kodama}\ and\ \citenamefont {Sasaki}(1984)}]{Kodama:1984ziu}%
  \BibitemOpen
  \bibfield  {author} {\bibinfo {author} {\bibfnamefont {H.}~\bibnamefont {Kodama}}\ and\ \bibinfo {author} {\bibfnamefont {M.}~\bibnamefont {Sasaki}},\ }\bibfield  {title} {\bibinfo {title} {{Cosmological Perturbation Theory}},\ }\href {https://doi.org/10.1143/PTPS.78.1} {\bibfield  {journal} {\bibinfo  {journal} {Prog. Theor. Phys. Suppl.}\ }\textbf {\bibinfo {volume} {78}},\ \bibinfo {pages} {1} (\bibinfo {year} {1984})}\BibitemShut {NoStop}%
\bibitem [{\citenamefont {Mukhanov}\ \emph {et~al.}(1992)\citenamefont {Mukhanov}, \citenamefont {Feldman},\ and\ \citenamefont {Brandenberger}}]{Mukhanov:1990me}%
  \BibitemOpen
  \bibfield  {author} {\bibinfo {author} {\bibfnamefont {V.~F.}\ \bibnamefont {Mukhanov}}, \bibinfo {author} {\bibfnamefont {H.~A.}\ \bibnamefont {Feldman}},\ and\ \bibinfo {author} {\bibfnamefont {R.~H.}\ \bibnamefont {Brandenberger}},\ }\bibfield  {title} {\bibinfo {title} {{Theory of cosmological perturbations. Part 1. Classical perturbations. Part 2. Quantum theory of perturbations. Part 3. Extensions}},\ }\href {https://doi.org/10.1016/0370-1573(92)90044-Z} {\bibfield  {journal} {\bibinfo  {journal} {Phys. Rept.}\ }\textbf {\bibinfo {volume} {215}},\ \bibinfo {pages} {203} (\bibinfo {year} {1992})}\BibitemShut {NoStop}%
\bibitem [{\citenamefont {Ma}\ and\ \citenamefont {Bertschinger}(1995)}]{Ma:1995ey}%
  \BibitemOpen
  \bibfield  {author} {\bibinfo {author} {\bibfnamefont {C.-P.}\ \bibnamefont {Ma}}\ and\ \bibinfo {author} {\bibfnamefont {E.}~\bibnamefont {Bertschinger}},\ }\bibfield  {title} {\bibinfo {title} {{Cosmological perturbation theory in the synchronous and conformal Newtonian gauges}},\ }\href {https://doi.org/10.1086/176550} {\bibfield  {journal} {\bibinfo  {journal} {Astrophys. J.}\ }\textbf {\bibinfo {volume} {455}},\ \bibinfo {pages} {7} (\bibinfo {year} {1995})},\ \Eprint {https://arxiv.org/abs/astro-ph/9506072} {arXiv:astro-ph/9506072} \BibitemShut {NoStop}%
\bibitem [{\citenamefont {Lyth}\ and\ \citenamefont {Riotto}(1999)}]{Lyth:1998xn}%
  \BibitemOpen
  \bibfield  {author} {\bibinfo {author} {\bibfnamefont {D.~H.}\ \bibnamefont {Lyth}}\ and\ \bibinfo {author} {\bibfnamefont {A.}~\bibnamefont {Riotto}},\ }\bibfield  {title} {\bibinfo {title} {{Particle physics models of inflation and the cosmological density perturbation}},\ }\href {https://doi.org/10.1016/S0370-1573(98)00128-8} {\bibfield  {journal} {\bibinfo  {journal} {Phys. Rept.}\ }\textbf {\bibinfo {volume} {314}},\ \bibinfo {pages} {1} (\bibinfo {year} {1999})},\ \Eprint {https://arxiv.org/abs/hep-ph/9807278} {arXiv:hep-ph/9807278} \BibitemShut {NoStop}%
\bibitem [{\citenamefont {Kibble}(1976)}]{Kibble:1976sj}%
  \BibitemOpen
  \bibfield  {author} {\bibinfo {author} {\bibfnamefont {T.~W.~B.}\ \bibnamefont {Kibble}},\ }\bibfield  {title} {\bibinfo {title} {{Topology of Cosmic Domains and Strings}},\ }\href {https://doi.org/10.1088/0305-4470/9/8/029} {\bibfield  {journal} {\bibinfo  {journal} {J. Phys. A}\ }\textbf {\bibinfo {volume} {9}},\ \bibinfo {pages} {1387} (\bibinfo {year} {1976})}\BibitemShut {NoStop}%
\bibitem [{\citenamefont {Copeland}\ and\ \citenamefont {Kibble}(2010)}]{Copeland:2009ga}%
  \BibitemOpen
  \bibfield  {author} {\bibinfo {author} {\bibfnamefont {E.~J.}\ \bibnamefont {Copeland}}\ and\ \bibinfo {author} {\bibfnamefont {T.~W.~B.}\ \bibnamefont {Kibble}},\ }\bibfield  {title} {\bibinfo {title} {{Cosmic Strings and Superstrings}},\ }\href {https://doi.org/10.1098/rspa.2009.0591} {\bibfield  {journal} {\bibinfo  {journal} {Proc. Roy. Soc. Lond. A}\ }\textbf {\bibinfo {volume} {466}},\ \bibinfo {pages} {623} (\bibinfo {year} {2010})},\ \Eprint {https://arxiv.org/abs/0911.1345} {arXiv:0911.1345 [hep-th]} \BibitemShut {NoStop}%
\bibitem [{\citenamefont {Zeldovich}\ \emph {et~al.}(1974)\citenamefont {Zeldovich}, \citenamefont {Kobzarev},\ and\ \citenamefont {Okun}}]{Zeldovich:1974uw}%
  \BibitemOpen
  \bibfield  {author} {\bibinfo {author} {\bibfnamefont {Y.~B.}\ \bibnamefont {Zeldovich}}, \bibinfo {author} {\bibfnamefont {I.~Y.}\ \bibnamefont {Kobzarev}},\ and\ \bibinfo {author} {\bibfnamefont {L.~B.}\ \bibnamefont {Okun}},\ }\bibfield  {title} {\bibinfo {title} {{Cosmological Consequences of the Spontaneous Breakdown of Discrete Symmetry}},\ }\href@noop {} {\bibfield  {journal} {\bibinfo  {journal} {Zh. Eksp. Teor. Fiz.}\ }\textbf {\bibinfo {volume} {67}},\ \bibinfo {pages} {3} (\bibinfo {year} {1974})}\BibitemShut {NoStop}%
\bibitem [{\citenamefont {Goldstone}\ \emph {et~al.}(1962)\citenamefont {Goldstone}, \citenamefont {Salam},\ and\ \citenamefont {Weinberg}}]{Goldstone:1962es}%
  \BibitemOpen
  \bibfield  {author} {\bibinfo {author} {\bibfnamefont {J.}~\bibnamefont {Goldstone}}, \bibinfo {author} {\bibfnamefont {A.}~\bibnamefont {Salam}},\ and\ \bibinfo {author} {\bibfnamefont {S.}~\bibnamefont {Weinberg}},\ }\bibfield  {title} {\bibinfo {title} {{Broken Symmetries}},\ }\href {https://doi.org/10.1103/PhysRev.127.965} {\bibfield  {journal} {\bibinfo  {journal} {Phys. Rev.}\ }\textbf {\bibinfo {volume} {127}},\ \bibinfo {pages} {965} (\bibinfo {year} {1962})}\BibitemShut {NoStop}%
\bibitem [{\citenamefont {Kirzhnits}(1972)}]{Kirzhnits:1972iw}%
  \BibitemOpen
  \bibfield  {author} {\bibinfo {author} {\bibfnamefont {D.~A.}\ \bibnamefont {Kirzhnits}},\ }\bibfield  {title} {\bibinfo {title} {{Weinberg model in the hot universe}},\ }\href@noop {} {\bibfield  {journal} {\bibinfo  {journal} {JETP Lett.}\ }\textbf {\bibinfo {volume} {15}},\ \bibinfo {pages} {529} (\bibinfo {year} {1972})}\BibitemShut {NoStop}%
\bibitem [{\citenamefont {Coleman}\ and\ \citenamefont {Weinberg}(1973)}]{Coleman:1973jx}%
  \BibitemOpen
  \bibfield  {author} {\bibinfo {author} {\bibfnamefont {S.~R.}\ \bibnamefont {Coleman}}\ and\ \bibinfo {author} {\bibfnamefont {E.~J.}\ \bibnamefont {Weinberg}},\ }\bibfield  {title} {\bibinfo {title} {{Radiative Corrections as the Origin of Spontaneous Symmetry Breaking}},\ }\href {https://doi.org/10.1103/PhysRevD.7.1888} {\bibfield  {journal} {\bibinfo  {journal} {Phys. Rev. D}\ }\textbf {\bibinfo {volume} {7}},\ \bibinfo {pages} {1888} (\bibinfo {year} {1973})}\BibitemShut {NoStop}%
\bibitem [{\citenamefont {Dolan}\ and\ \citenamefont {Jackiw}(1974)}]{Dolan:1973qd}%
  \BibitemOpen
  \bibfield  {author} {\bibinfo {author} {\bibfnamefont {L.}~\bibnamefont {Dolan}}\ and\ \bibinfo {author} {\bibfnamefont {R.}~\bibnamefont {Jackiw}},\ }\bibfield  {title} {\bibinfo {title} {{Symmetry Behavior at Finite Temperature}},\ }\href {https://doi.org/10.1103/PhysRevD.9.3320} {\bibfield  {journal} {\bibinfo  {journal} {Phys. Rev. D}\ }\textbf {\bibinfo {volume} {9}},\ \bibinfo {pages} {3320} (\bibinfo {year} {1974})}\BibitemShut {NoStop}%
\bibitem [{\citenamefont {Kirzhnits}\ and\ \citenamefont {Linde}(1976)}]{Kirzhnits:1976ts}%
  \BibitemOpen
  \bibfield  {author} {\bibinfo {author} {\bibfnamefont {D.~A.}\ \bibnamefont {Kirzhnits}}\ and\ \bibinfo {author} {\bibfnamefont {A.~D.}\ \bibnamefont {Linde}},\ }\bibfield  {title} {\bibinfo {title} {{Symmetry Behavior in Gauge Theories}},\ }\href {https://doi.org/10.1016/0003-4916(76)90279-7} {\bibfield  {journal} {\bibinfo  {journal} {Annals Phys.}\ }\textbf {\bibinfo {volume} {101}},\ \bibinfo {pages} {195} (\bibinfo {year} {1976})}\BibitemShut {NoStop}%
\bibitem [{\citenamefont {Weinberg}(1974)}]{Weinberg:1974hy}%
  \BibitemOpen
  \bibfield  {author} {\bibinfo {author} {\bibfnamefont {S.}~\bibnamefont {Weinberg}},\ }\bibfield  {title} {\bibinfo {title} {{Gauge and Global Symmetries at High Temperature}},\ }\href {https://doi.org/10.1103/PhysRevD.9.3357} {\bibfield  {journal} {\bibinfo  {journal} {Phys. Rev. D}\ }\textbf {\bibinfo {volume} {9}},\ \bibinfo {pages} {3357} (\bibinfo {year} {1974})}\BibitemShut {NoStop}%
\bibitem [{\citenamefont {Witten}(1984)}]{Witten:1984rs}%
  \BibitemOpen
  \bibfield  {author} {\bibinfo {author} {\bibfnamefont {E.}~\bibnamefont {Witten}},\ }\bibfield  {title} {\bibinfo {title} {{Cosmic Separation of Phases}},\ }\href {https://doi.org/10.1103/PhysRevD.30.272} {\bibfield  {journal} {\bibinfo  {journal} {Phys. Rev. D}\ }\textbf {\bibinfo {volume} {30}},\ \bibinfo {pages} {272} (\bibinfo {year} {1984})}\BibitemShut {NoStop}%
\bibitem [{\citenamefont {Hogan}(1986)}]{hogan1986gravitational}%
  \BibitemOpen
  \bibfield  {author} {\bibinfo {author} {\bibfnamefont {C.}~\bibnamefont {Hogan}},\ }\bibfield  {title} {\bibinfo {title} {Gravitational radiation from cosmological phase transitions},\ }\href@noop {} {\bibfield  {journal} {\bibinfo  {journal} {Monthly Notices of the Royal Astronomical Society}\ }\textbf {\bibinfo {volume} {218}},\ \bibinfo {pages} {629} (\bibinfo {year} {1986})}\BibitemShut {NoStop}%
\bibitem [{\citenamefont {Carrington}(1992)}]{Carrington:1991hz}%
  \BibitemOpen
  \bibfield  {author} {\bibinfo {author} {\bibfnamefont {M.~E.}\ \bibnamefont {Carrington}},\ }\bibfield  {title} {\bibinfo {title} {{The Effective potential at finite temperature in the Standard Model}},\ }\href {https://doi.org/10.1103/PhysRevD.45.2933} {\bibfield  {journal} {\bibinfo  {journal} {Phys. Rev. D}\ }\textbf {\bibinfo {volume} {45}},\ \bibinfo {pages} {2933} (\bibinfo {year} {1992})}\BibitemShut {NoStop}%
\bibitem [{\citenamefont {Arnold}(1992)}]{Arnold:1992fb}%
  \BibitemOpen
  \bibfield  {author} {\bibinfo {author} {\bibfnamefont {P.~B.}\ \bibnamefont {Arnold}},\ }\bibfield  {title} {\bibinfo {title} {{Phase transition temperatures at next-to-leading order}},\ }\href {https://doi.org/10.1103/PhysRevD.46.2628} {\bibfield  {journal} {\bibinfo  {journal} {Phys. Rev. D}\ }\textbf {\bibinfo {volume} {46}},\ \bibinfo {pages} {2628} (\bibinfo {year} {1992})},\ \Eprint {https://arxiv.org/abs/hep-ph/9204228} {arXiv:hep-ph/9204228} \BibitemShut {NoStop}%
\bibitem [{\citenamefont {Hindmarsh}\ \emph {et~al.}(2021)\citenamefont {Hindmarsh}, \citenamefont {L\"uben}, \citenamefont {Lumma},\ and\ \citenamefont {Pauly}}]{Hindmarsh:2020hop}%
  \BibitemOpen
  \bibfield  {author} {\bibinfo {author} {\bibfnamefont {M.~B.}\ \bibnamefont {Hindmarsh}}, \bibinfo {author} {\bibfnamefont {M.}~\bibnamefont {L\"uben}}, \bibinfo {author} {\bibfnamefont {J.}~\bibnamefont {Lumma}},\ and\ \bibinfo {author} {\bibfnamefont {M.}~\bibnamefont {Pauly}},\ }\bibfield  {title} {\bibinfo {title} {{Phase transitions in the early universe}},\ }\href {https://doi.org/10.21468/SciPostPhysLectNotes.24} {\bibfield  {journal} {\bibinfo  {journal} {SciPost Phys. Lect. Notes}\ }\textbf {\bibinfo {volume} {24}},\ \bibinfo {pages} {1} (\bibinfo {year} {2021})},\ \Eprint {https://arxiv.org/abs/2008.09136} {arXiv:2008.09136 [astro-ph.CO]} \BibitemShut {NoStop}%
\bibitem [{\citenamefont {Dom\`enech}\ \emph {et~al.}(2021)\citenamefont {Dom\`enech}, \citenamefont {Lin},\ and\ \citenamefont {Sasaki}}]{Domenech:2020ssp}%
  \BibitemOpen
  \bibfield  {author} {\bibinfo {author} {\bibfnamefont {G.}~\bibnamefont {Dom\`enech}}, \bibinfo {author} {\bibfnamefont {C.}~\bibnamefont {Lin}},\ and\ \bibinfo {author} {\bibfnamefont {M.}~\bibnamefont {Sasaki}},\ }\bibfield  {title} {\bibinfo {title} {{Gravitational wave constraints on the primordial black hole dominated early universe}},\ }\href {https://doi.org/10.1088/1475-7516/2021/11/E01} {\bibfield  {journal} {\bibinfo  {journal} {JCAP}\ }\textbf {\bibinfo {volume} {04}},\ \bibinfo {pages} {062}},\ \bibinfo {note} {[Erratum: JCAP 11, E01 (2021)]},\ \Eprint {https://arxiv.org/abs/2012.08151} {arXiv:2012.08151 [gr-qc]} \BibitemShut {NoStop}%
\bibitem [{\citenamefont {Athron}\ \emph {et~al.}(2024)\citenamefont {Athron}, \citenamefont {Bal\'azs}, \citenamefont {Fowlie}, \citenamefont {Morris},\ and\ \citenamefont {Wu}}]{Athron:2023xlk}%
  \BibitemOpen
  \bibfield  {author} {\bibinfo {author} {\bibfnamefont {P.}~\bibnamefont {Athron}}, \bibinfo {author} {\bibfnamefont {C.}~\bibnamefont {Bal\'azs}}, \bibinfo {author} {\bibfnamefont {A.}~\bibnamefont {Fowlie}}, \bibinfo {author} {\bibfnamefont {L.}~\bibnamefont {Morris}},\ and\ \bibinfo {author} {\bibfnamefont {L.}~\bibnamefont {Wu}},\ }\bibfield  {title} {\bibinfo {title} {{Cosmological phase transitions: From perturbative particle physics to gravitational waves}},\ }\href {https://doi.org/10.1016/j.ppnp.2023.104094} {\bibfield  {journal} {\bibinfo  {journal} {Prog. Part. Nucl. Phys.}\ }\textbf {\bibinfo {volume} {135}},\ \bibinfo {pages} {104094} (\bibinfo {year} {2024})},\ \Eprint {https://arxiv.org/abs/2305.02357} {arXiv:2305.02357 [hep-ph]} \BibitemShut {NoStop}%
\bibitem [{\citenamefont {Dom\`enech}\ and\ \citenamefont {Tr\"ankle}(2024)}]{Domenech:2024wao}%
  \BibitemOpen
  \bibfield  {author} {\bibinfo {author} {\bibfnamefont {G.}~\bibnamefont {Dom\`enech}}\ and\ \bibinfo {author} {\bibfnamefont {J.}~\bibnamefont {Tr\"ankle}},\ }\href@noop {} {\bibinfo {title} {{From formation to evaporation: Induced gravitational wave probes of the primordial black hole reheating scenario}}} (\bibinfo {year} {2024}),\ \Eprint {https://arxiv.org/abs/2409.12125} {arXiv:2409.12125 [gr-qc]} \BibitemShut {NoStop}%
\bibitem [{\citenamefont {Roshan}\ and\ \citenamefont {White}(2024)}]{Roshan:2024qnv}%
  \BibitemOpen
  \bibfield  {author} {\bibinfo {author} {\bibfnamefont {R.}~\bibnamefont {Roshan}}\ and\ \bibinfo {author} {\bibfnamefont {G.}~\bibnamefont {White}},\ }\href@noop {} {\bibinfo {title} {{Using gravitational waves to see the first second of the Universe}}} (\bibinfo {year} {2024}),\ \Eprint {https://arxiv.org/abs/2401.04388} {arXiv:2401.04388 [hep-ph]} \BibitemShut {NoStop}%
\bibitem [{\citenamefont {Kamionkowski}\ \emph {et~al.}(1994)\citenamefont {Kamionkowski}, \citenamefont {Kosowsky},\ and\ \citenamefont {Turner}}]{Kamionkowski:1993fg}%
  \BibitemOpen
  \bibfield  {author} {\bibinfo {author} {\bibfnamefont {M.}~\bibnamefont {Kamionkowski}}, \bibinfo {author} {\bibfnamefont {A.}~\bibnamefont {Kosowsky}},\ and\ \bibinfo {author} {\bibfnamefont {M.~S.}\ \bibnamefont {Turner}},\ }\bibfield  {title} {\bibinfo {title} {{Gravitational radiation from first order phase transitions}},\ }\href {https://doi.org/10.1103/PhysRevD.49.2837} {\bibfield  {journal} {\bibinfo  {journal} {Phys. Rev. D}\ }\textbf {\bibinfo {volume} {49}},\ \bibinfo {pages} {2837} (\bibinfo {year} {1994})},\ \Eprint {https://arxiv.org/abs/astro-ph/9310044} {arXiv:astro-ph/9310044} \BibitemShut {NoStop}%
\bibitem [{\citenamefont {Kosowsky}\ \emph {et~al.}(1992)\citenamefont {Kosowsky}, \citenamefont {Turner},\ and\ \citenamefont {Watkins}}]{Kosowsky:1992rz}%
  \BibitemOpen
  \bibfield  {author} {\bibinfo {author} {\bibfnamefont {A.}~\bibnamefont {Kosowsky}}, \bibinfo {author} {\bibfnamefont {M.~S.}\ \bibnamefont {Turner}},\ and\ \bibinfo {author} {\bibfnamefont {R.}~\bibnamefont {Watkins}},\ }\bibfield  {title} {\bibinfo {title} {{Gravitational waves from first order cosmological phase transitions}},\ }\href {https://doi.org/10.1103/PhysRevLett.69.2026} {\bibfield  {journal} {\bibinfo  {journal} {Phys. Rev. Lett.}\ }\textbf {\bibinfo {volume} {69}},\ \bibinfo {pages} {2026} (\bibinfo {year} {1992})}\BibitemShut {NoStop}%
\bibitem [{\citenamefont {Kosowsky}\ and\ \citenamefont {Turner}(1993)}]{Kosowsky:1992vn}%
  \BibitemOpen
  \bibfield  {author} {\bibinfo {author} {\bibfnamefont {A.}~\bibnamefont {Kosowsky}}\ and\ \bibinfo {author} {\bibfnamefont {M.~S.}\ \bibnamefont {Turner}},\ }\bibfield  {title} {\bibinfo {title} {{Gravitational radiation from colliding vacuum bubbles: envelope approximation to many bubble collisions}},\ }\href {https://doi.org/10.1103/PhysRevD.47.4372} {\bibfield  {journal} {\bibinfo  {journal} {Phys. Rev. D}\ }\textbf {\bibinfo {volume} {47}},\ \bibinfo {pages} {4372} (\bibinfo {year} {1993})},\ \Eprint {https://arxiv.org/abs/astro-ph/9211004} {arXiv:astro-ph/9211004} \BibitemShut {NoStop}%
\bibitem [{\citenamefont {Kosowsky}\ \emph {et~al.}(2002)\citenamefont {Kosowsky}, \citenamefont {Mack},\ and\ \citenamefont {Kahniashvili}}]{Kosowsky:2001xp}%
  \BibitemOpen
  \bibfield  {author} {\bibinfo {author} {\bibfnamefont {A.}~\bibnamefont {Kosowsky}}, \bibinfo {author} {\bibfnamefont {A.}~\bibnamefont {Mack}},\ and\ \bibinfo {author} {\bibfnamefont {T.}~\bibnamefont {Kahniashvili}},\ }\bibfield  {title} {\bibinfo {title} {{Gravitational radiation from cosmological turbulence}},\ }\href {https://doi.org/10.1103/PhysRevD.66.024030} {\bibfield  {journal} {\bibinfo  {journal} {Phys. Rev. D}\ }\textbf {\bibinfo {volume} {66}},\ \bibinfo {pages} {024030} (\bibinfo {year} {2002})},\ \Eprint {https://arxiv.org/abs/astro-ph/0111483} {arXiv:astro-ph/0111483} \BibitemShut {NoStop}%
\bibitem [{\citenamefont {Davoudiasl}\ \emph {et~al.}(2005)\citenamefont {Davoudiasl}, \citenamefont {Kitano}, \citenamefont {Li},\ and\ \citenamefont {Murayama}}]{Davoudiasl:2004be}%
  \BibitemOpen
  \bibfield  {author} {\bibinfo {author} {\bibfnamefont {H.}~\bibnamefont {Davoudiasl}}, \bibinfo {author} {\bibfnamefont {R.}~\bibnamefont {Kitano}}, \bibinfo {author} {\bibfnamefont {T.}~\bibnamefont {Li}},\ and\ \bibinfo {author} {\bibfnamefont {H.}~\bibnamefont {Murayama}},\ }\bibfield  {title} {\bibinfo {title} {{The New minimal standard model}},\ }\href {https://doi.org/10.1016/j.physletb.2005.01.026} {\bibfield  {journal} {\bibinfo  {journal} {Phys. Lett. B}\ }\textbf {\bibinfo {volume} {609}},\ \bibinfo {pages} {117} (\bibinfo {year} {2005})},\ \Eprint {https://arxiv.org/abs/hep-ph/0405097} {arXiv:hep-ph/0405097} \BibitemShut {NoStop}%
\bibitem [{\citenamefont {Espinosa}\ and\ \citenamefont {Quiros}(2007)}]{Espinosa:2007qk}%
  \BibitemOpen
  \bibfield  {author} {\bibinfo {author} {\bibfnamefont {J.~R.}\ \bibnamefont {Espinosa}}\ and\ \bibinfo {author} {\bibfnamefont {M.}~\bibnamefont {Quiros}},\ }\bibfield  {title} {\bibinfo {title} {{Novel Effects in Electroweak Breaking from a Hidden Sector}},\ }\href {https://doi.org/10.1103/PhysRevD.76.076004} {\bibfield  {journal} {\bibinfo  {journal} {Phys. Rev. D}\ }\textbf {\bibinfo {volume} {76}},\ \bibinfo {pages} {076004} (\bibinfo {year} {2007})},\ \Eprint {https://arxiv.org/abs/hep-ph/0701145} {arXiv:hep-ph/0701145} \BibitemShut {NoStop}%
\bibitem [{\citenamefont {Espinosa}\ \emph {et~al.}(2008)\citenamefont {Espinosa}, \citenamefont {Konstandin}, \citenamefont {No},\ and\ \citenamefont {Quiros}}]{Espinosa:2008kw}%
  \BibitemOpen
  \bibfield  {author} {\bibinfo {author} {\bibfnamefont {J.~R.}\ \bibnamefont {Espinosa}}, \bibinfo {author} {\bibfnamefont {T.}~\bibnamefont {Konstandin}}, \bibinfo {author} {\bibfnamefont {J.~M.}\ \bibnamefont {No}},\ and\ \bibinfo {author} {\bibfnamefont {M.}~\bibnamefont {Quiros}},\ }\bibfield  {title} {\bibinfo {title} {{Some Cosmological Implications of Hidden Sectors}},\ }\href {https://doi.org/10.1103/PhysRevD.78.123528} {\bibfield  {journal} {\bibinfo  {journal} {Phys. Rev. D}\ }\textbf {\bibinfo {volume} {78}},\ \bibinfo {pages} {123528} (\bibinfo {year} {2008})},\ \Eprint {https://arxiv.org/abs/0809.3215} {arXiv:0809.3215 [hep-ph]} \BibitemShut {NoStop}%
\bibitem [{\citenamefont {Breitbach}\ \emph {et~al.}(2019)\citenamefont {Breitbach}, \citenamefont {Kopp}, \citenamefont {Madge}, \citenamefont {Opferkuch},\ and\ \citenamefont {Schwaller}}]{Breitbach:2018ddu}%
  \BibitemOpen
  \bibfield  {author} {\bibinfo {author} {\bibfnamefont {M.}~\bibnamefont {Breitbach}}, \bibinfo {author} {\bibfnamefont {J.}~\bibnamefont {Kopp}}, \bibinfo {author} {\bibfnamefont {E.}~\bibnamefont {Madge}}, \bibinfo {author} {\bibfnamefont {T.}~\bibnamefont {Opferkuch}},\ and\ \bibinfo {author} {\bibfnamefont {P.}~\bibnamefont {Schwaller}},\ }\bibfield  {title} {\bibinfo {title} {{Dark, Cold, and Noisy: Constraining Secluded Hidden Sectors with Gravitational Waves}},\ }\href {https://doi.org/10.1088/1475-7516/2019/07/007} {\bibfield  {journal} {\bibinfo  {journal} {JCAP}\ }\textbf {\bibinfo {volume} {07}},\ \bibinfo {pages} {007}},\ \Eprint {https://arxiv.org/abs/1811.11175} {arXiv:1811.11175 [hep-ph]} \BibitemShut {NoStop}%
\bibitem [{\citenamefont {Fairbairn}\ \emph {et~al.}(2019)\citenamefont {Fairbairn}, \citenamefont {Hardy},\ and\ \citenamefont {Wickens}}]{Fairbairn:2019xog}%
  \BibitemOpen
  \bibfield  {author} {\bibinfo {author} {\bibfnamefont {M.}~\bibnamefont {Fairbairn}}, \bibinfo {author} {\bibfnamefont {E.}~\bibnamefont {Hardy}},\ and\ \bibinfo {author} {\bibfnamefont {A.}~\bibnamefont {Wickens}},\ }\bibfield  {title} {\bibinfo {title} {{Hearing without seeing: gravitational waves from hot and cold hidden sectors}},\ }\href {https://doi.org/10.1007/JHEP07(2019)044} {\bibfield  {journal} {\bibinfo  {journal} {JHEP}\ }\textbf {\bibinfo {volume} {07}},\ \bibinfo {pages} {044}},\ \Eprint {https://arxiv.org/abs/1901.11038} {arXiv:1901.11038 [hep-ph]} \BibitemShut {NoStop}%
\bibitem [{\citenamefont {Adams}(1993)}]{Adams:1993zs}%
  \BibitemOpen
  \bibfield  {author} {\bibinfo {author} {\bibfnamefont {F.~C.}\ \bibnamefont {Adams}},\ }\bibfield  {title} {\bibinfo {title} {{General solutions for tunneling of scalar fields with quartic potentials}},\ }\href {https://doi.org/10.1103/PhysRevD.48.2800} {\bibfield  {journal} {\bibinfo  {journal} {Phys. Rev. D}\ }\textbf {\bibinfo {volume} {48}},\ \bibinfo {pages} {2800} (\bibinfo {year} {1993})},\ \Eprint {https://arxiv.org/abs/hep-ph/9302321} {arXiv:hep-ph/9302321} \BibitemShut {NoStop}%
\bibitem [{\citenamefont {Dine}\ \emph {et~al.}(1992)\citenamefont {Dine}, \citenamefont {Leigh}, \citenamefont {Huet}, \citenamefont {Linde},\ and\ \citenamefont {Linde}}]{Dine:1992wr}%
  \BibitemOpen
  \bibfield  {author} {\bibinfo {author} {\bibfnamefont {M.}~\bibnamefont {Dine}}, \bibinfo {author} {\bibfnamefont {R.~G.}\ \bibnamefont {Leigh}}, \bibinfo {author} {\bibfnamefont {P.~Y.}\ \bibnamefont {Huet}}, \bibinfo {author} {\bibfnamefont {A.~D.}\ \bibnamefont {Linde}},\ and\ \bibinfo {author} {\bibfnamefont {D.~A.}\ \bibnamefont {Linde}},\ }\bibfield  {title} {\bibinfo {title} {{Towards the theory of the electroweak phase transition}},\ }\href {https://doi.org/10.1103/PhysRevD.46.550} {\bibfield  {journal} {\bibinfo  {journal} {Phys. Rev. D}\ }\textbf {\bibinfo {volume} {46}},\ \bibinfo {pages} {550} (\bibinfo {year} {1992})},\ \Eprint {https://arxiv.org/abs/hep-ph/9203203} {arXiv:hep-ph/9203203} \BibitemShut {NoStop}%
\bibitem [{\citenamefont {Ellis}\ \emph {et~al.}(2019)\citenamefont {Ellis}, \citenamefont {Lewicki}, \citenamefont {No},\ and\ \citenamefont {Vaskonen}}]{Ellis:2019oqb}%
  \BibitemOpen
  \bibfield  {author} {\bibinfo {author} {\bibfnamefont {J.}~\bibnamefont {Ellis}}, \bibinfo {author} {\bibfnamefont {M.}~\bibnamefont {Lewicki}}, \bibinfo {author} {\bibfnamefont {J.~M.}\ \bibnamefont {No}},\ and\ \bibinfo {author} {\bibfnamefont {V.}~\bibnamefont {Vaskonen}},\ }\bibfield  {title} {\bibinfo {title} {{Gravitational wave energy budget in strongly supercooled phase transitions}},\ }\href {https://doi.org/10.1088/1475-7516/2019/06/024} {\bibfield  {journal} {\bibinfo  {journal} {JCAP}\ }\textbf {\bibinfo {volume} {06}},\ \bibinfo {pages} {024}},\ \Eprint {https://arxiv.org/abs/1903.09642} {arXiv:1903.09642 [hep-ph]} \BibitemShut {NoStop}%
\bibitem [{\citenamefont {Aghanim}\ \emph {et~al.}(2020)\citenamefont {Aghanim} \emph {et~al.}}]{Planck:2018vyg}%
  \BibitemOpen
  \bibfield  {author} {\bibinfo {author} {\bibfnamefont {N.}~\bibnamefont {Aghanim}} \emph {et~al.} (\bibinfo {collaboration} {Planck}),\ }\bibfield  {title} {\bibinfo {title} {{Planck 2018 results. VI. Cosmological parameters}},\ }\href {https://doi.org/10.1051/0004-6361/201833910} {\bibfield  {journal} {\bibinfo  {journal} {Astron. Astrophys.}\ }\textbf {\bibinfo {volume} {641}},\ \bibinfo {pages} {A6} (\bibinfo {year} {2020})},\ \bibinfo {note} {[Erratum: Astron.Astrophys. 652, C4 (2021)]},\ \Eprint {https://arxiv.org/abs/1807.06209} {arXiv:1807.06209 [astro-ph.CO]} \BibitemShut {NoStop}%
\bibitem [{\citenamefont {Espinosa}\ \emph {et~al.}(2010)\citenamefont {Espinosa}, \citenamefont {Konstandin}, \citenamefont {No},\ and\ \citenamefont {Servant}}]{Espinosa:2010hh}%
  \BibitemOpen
  \bibfield  {author} {\bibinfo {author} {\bibfnamefont {J.~R.}\ \bibnamefont {Espinosa}}, \bibinfo {author} {\bibfnamefont {T.}~\bibnamefont {Konstandin}}, \bibinfo {author} {\bibfnamefont {J.~M.}\ \bibnamefont {No}},\ and\ \bibinfo {author} {\bibfnamefont {G.}~\bibnamefont {Servant}},\ }\bibfield  {title} {\bibinfo {title} {{Energy Budget of Cosmological First-order Phase Transitions}},\ }\href {https://doi.org/10.1088/1475-7516/2010/06/028} {\bibfield  {journal} {\bibinfo  {journal} {JCAP}\ }\textbf {\bibinfo {volume} {06}},\ \bibinfo {pages} {028}},\ \Eprint {https://arxiv.org/abs/1004.4187} {arXiv:1004.4187 [hep-ph]} \BibitemShut {NoStop}%
\bibitem [{\citenamefont {Barni}\ \emph {et~al.}(2024)\citenamefont {Barni}, \citenamefont {Blasi},\ and\ \citenamefont {Vanvlasselaer}}]{Barni:2024lkj}%
  \BibitemOpen
  \bibfield  {author} {\bibinfo {author} {\bibfnamefont {G.}~\bibnamefont {Barni}}, \bibinfo {author} {\bibfnamefont {S.}~\bibnamefont {Blasi}},\ and\ \bibinfo {author} {\bibfnamefont {M.}~\bibnamefont {Vanvlasselaer}},\ }\bibfield  {title} {\bibinfo {title} {{The hydrodynamics of inverse phase transitions}},\ }\href {https://doi.org/10.1088/1475-7516/2024/10/042} {\bibfield  {journal} {\bibinfo  {journal} {JCAP}\ }\textbf {\bibinfo {volume} {10}},\ \bibinfo {pages} {042}},\ \Eprint {https://arxiv.org/abs/2406.01596} {arXiv:2406.01596 [hep-ph]} \BibitemShut {NoStop}%
\bibitem [{\citenamefont {Azatov}\ \emph {et~al.}(2024{\natexlab{a}})\citenamefont {Azatov}, \citenamefont {Barni},\ and\ \citenamefont {Petrossian-Byrne}}]{Azatov:2024auq}%
  \BibitemOpen
  \bibfield  {author} {\bibinfo {author} {\bibfnamefont {A.}~\bibnamefont {Azatov}}, \bibinfo {author} {\bibfnamefont {G.}~\bibnamefont {Barni}},\ and\ \bibinfo {author} {\bibfnamefont {R.}~\bibnamefont {Petrossian-Byrne}},\ }\href@noop {} {\bibinfo {title} {Nlo friction in symmetry restoring phase transitions}} (\bibinfo {year} {2024}{\natexlab{a}}),\ \bibinfo {note} {preprint},\ \Eprint {https://arxiv.org/abs/2405.19447} {arXiv:2405.19447 [hep-ph]} \BibitemShut {NoStop}%
\bibitem [{\citenamefont {Huber}\ and\ \citenamefont {Konstandin}(2008)}]{Huber:2008hg}%
  \BibitemOpen
  \bibfield  {author} {\bibinfo {author} {\bibfnamefont {S.~J.}\ \bibnamefont {Huber}}\ and\ \bibinfo {author} {\bibfnamefont {T.}~\bibnamefont {Konstandin}},\ }\bibfield  {title} {\bibinfo {title} {{Gravitational Wave Production by Collisions: More Bubbles}},\ }\href {https://doi.org/10.1088/1475-7516/2008/09/022} {\bibfield  {journal} {\bibinfo  {journal} {JCAP}\ }\textbf {\bibinfo {volume} {09}},\ \bibinfo {pages} {022}},\ \Eprint {https://arxiv.org/abs/0806.1828} {arXiv:0806.1828 [hep-ph]} \BibitemShut {NoStop}%
\bibitem [{\citenamefont {Caprini}\ \emph {et~al.}(2009)\citenamefont {Caprini}, \citenamefont {Durrer},\ and\ \citenamefont {Servant}}]{Caprini:2009yp}%
  \BibitemOpen
  \bibfield  {author} {\bibinfo {author} {\bibfnamefont {C.}~\bibnamefont {Caprini}}, \bibinfo {author} {\bibfnamefont {R.}~\bibnamefont {Durrer}},\ and\ \bibinfo {author} {\bibfnamefont {G.}~\bibnamefont {Servant}},\ }\bibfield  {title} {\bibinfo {title} {{The stochastic gravitational wave background from turbulence and magnetic fields generated by a first-order phase transition}},\ }\href {https://doi.org/10.1088/1475-7516/2009/12/024} {\bibfield  {journal} {\bibinfo  {journal} {JCAP}\ }\textbf {\bibinfo {volume} {12}},\ \bibinfo {pages} {024}},\ \Eprint {https://arxiv.org/abs/0909.0622} {arXiv:0909.0622 [astro-ph.CO]} \BibitemShut {NoStop}%
\bibitem [{\citenamefont {Ellis}\ \emph {et~al.}(2020)\citenamefont {Ellis}, \citenamefont {Lewicki},\ and\ \citenamefont {No}}]{Ellis:2020awk}%
  \BibitemOpen
  \bibfield  {author} {\bibinfo {author} {\bibfnamefont {J.}~\bibnamefont {Ellis}}, \bibinfo {author} {\bibfnamefont {M.}~\bibnamefont {Lewicki}},\ and\ \bibinfo {author} {\bibfnamefont {J.~M.}\ \bibnamefont {No}},\ }\bibfield  {title} {\bibinfo {title} {{Gravitational waves from first-order cosmological phase transitions: lifetime of the sound wave source}},\ }\href {https://doi.org/10.1088/1475-7516/2020/07/050} {\bibfield  {journal} {\bibinfo  {journal} {JCAP}\ }\textbf {\bibinfo {volume} {07}},\ \bibinfo {pages} {050}},\ \Eprint {https://arxiv.org/abs/2003.07360} {arXiv:2003.07360 [hep-ph]} \BibitemShut {NoStop}%
\bibitem [{\citenamefont {Caprini}\ \emph {et~al.}(2020)\citenamefont {Caprini} \emph {et~al.}}]{Caprini:2019egz}%
  \BibitemOpen
  \bibfield  {author} {\bibinfo {author} {\bibfnamefont {C.}~\bibnamefont {Caprini}} \emph {et~al.},\ }\bibfield  {title} {\bibinfo {title} {{Detecting gravitational waves from cosmological phase transitions with LISA: an update}},\ }\href {https://doi.org/10.1088/1475-7516/2020/03/024} {\bibfield  {journal} {\bibinfo  {journal} {JCAP}\ }\textbf {\bibinfo {volume} {03}},\ \bibinfo {pages} {024}},\ \Eprint {https://arxiv.org/abs/1910.13125} {arXiv:1910.13125 [astro-ph.CO]} \BibitemShut {NoStop}%
\bibitem [{\citenamefont {Crowder}\ and\ \citenamefont {Cornish}(2005)}]{Crowder:2005nr}%
  \BibitemOpen
  \bibfield  {author} {\bibinfo {author} {\bibfnamefont {J.}~\bibnamefont {Crowder}}\ and\ \bibinfo {author} {\bibfnamefont {N.~J.}\ \bibnamefont {Cornish}},\ }\bibfield  {title} {\bibinfo {title} {{Beyond LISA: Exploring future gravitational wave missions}},\ }\href {https://doi.org/10.1103/PhysRevD.72.083005} {\bibfield  {journal} {\bibinfo  {journal} {Phys. Rev. D}\ }\textbf {\bibinfo {volume} {72}},\ \bibinfo {pages} {083005} (\bibinfo {year} {2005})},\ \Eprint {https://arxiv.org/abs/gr-qc/0506015} {arXiv:gr-qc/0506015} \BibitemShut {NoStop}%
\bibitem [{\citenamefont {Braglia}\ and\ \citenamefont {Kuroyanagi}(2021)}]{Braglia:2021fxn}%
  \BibitemOpen
  \bibfield  {author} {\bibinfo {author} {\bibfnamefont {M.}~\bibnamefont {Braglia}}\ and\ \bibinfo {author} {\bibfnamefont {S.}~\bibnamefont {Kuroyanagi}},\ }\bibfield  {title} {\bibinfo {title} {{Probing prerecombination physics by the cross-correlation of stochastic gravitational waves and CMB anisotropies}},\ }\href {https://doi.org/10.1103/PhysRevD.104.123547} {\bibfield  {journal} {\bibinfo  {journal} {Phys. Rev. D}\ }\textbf {\bibinfo {volume} {104}},\ \bibinfo {pages} {123547} (\bibinfo {year} {2021})},\ \Eprint {https://arxiv.org/abs/2106.03786} {arXiv:2106.03786 [astro-ph.CO]} \BibitemShut {NoStop}%
\bibitem [{\citenamefont {Sesana}\ \emph {et~al.}(2021)\citenamefont {Sesana} \emph {et~al.}}]{Sesana:2019vho}%
  \BibitemOpen
  \bibfield  {author} {\bibinfo {author} {\bibfnamefont {A.}~\bibnamefont {Sesana}} \emph {et~al.},\ }\bibfield  {title} {\bibinfo {title} {{Unveiling the gravitational universe at $\mu$-Hz frequencies}},\ }\href {https://doi.org/10.1007/s10686-021-09709-9} {\bibfield  {journal} {\bibinfo  {journal} {Exper. Astron.}\ }\textbf {\bibinfo {volume} {51}},\ \bibinfo {pages} {1333} (\bibinfo {year} {2021})},\ \Eprint {https://arxiv.org/abs/1908.11391} {arXiv:1908.11391 [astro-ph.IM]} \BibitemShut {NoStop}%
\bibitem [{\citenamefont {Robson}\ \emph {et~al.}(2019)\citenamefont {Robson}, \citenamefont {Cornish},\ and\ \citenamefont {Liu}}]{Robson:2018ifk}%
  \BibitemOpen
  \bibfield  {author} {\bibinfo {author} {\bibfnamefont {T.}~\bibnamefont {Robson}}, \bibinfo {author} {\bibfnamefont {N.~J.}\ \bibnamefont {Cornish}},\ and\ \bibinfo {author} {\bibfnamefont {C.}~\bibnamefont {Liu}},\ }\bibfield  {title} {\bibinfo {title} {{The construction and use of LISA sensitivity curves}},\ }\href {https://doi.org/10.1088/1361-6382/ab1101} {\bibfield  {journal} {\bibinfo  {journal} {Class. Quant. Grav.}\ }\textbf {\bibinfo {volume} {36}},\ \bibinfo {pages} {105011} (\bibinfo {year} {2019})},\ \Eprint {https://arxiv.org/abs/1803.01944} {arXiv:1803.01944 [astro-ph.HE]} \BibitemShut {NoStop}%
\bibitem [{\citenamefont {Farmer}\ and\ \citenamefont {Phinney}(2003)}]{Farmer:2003pa}%
  \BibitemOpen
  \bibfield  {author} {\bibinfo {author} {\bibfnamefont {A.~J.}\ \bibnamefont {Farmer}}\ and\ \bibinfo {author} {\bibfnamefont {E.~S.}\ \bibnamefont {Phinney}},\ }\bibfield  {title} {\bibinfo {title} {{The gravitational wave background from cosmological compact binaries}},\ }\href {https://doi.org/10.1111/j.1365-2966.2003.07176.x} {\bibfield  {journal} {\bibinfo  {journal} {Mon. Not. Roy. Astron. Soc.}\ }\textbf {\bibinfo {volume} {346}},\ \bibinfo {pages} {1197} (\bibinfo {year} {2003})},\ \Eprint {https://arxiv.org/abs/astro-ph/0304393} {arXiv:astro-ph/0304393} \BibitemShut {NoStop}%
\bibitem [{\citenamefont {Guo}\ \emph {et~al.}(2024)\citenamefont {Guo}, \citenamefont {Hajkarim}, \citenamefont {Sinha}, \citenamefont {White},\ and\ \citenamefont {Xiao}}]{guo2024precise}%
  \BibitemOpen
  \bibfield  {author} {\bibinfo {author} {\bibfnamefont {H.-k.}\ \bibnamefont {Guo}}, \bibinfo {author} {\bibfnamefont {F.}~\bibnamefont {Hajkarim}}, \bibinfo {author} {\bibfnamefont {K.}~\bibnamefont {Sinha}}, \bibinfo {author} {\bibfnamefont {G.}~\bibnamefont {White}},\ and\ \bibinfo {author} {\bibfnamefont {Y.}~\bibnamefont {Xiao}},\ }\bibfield  {title} {\bibinfo {title} {A precise fitting formula for gravitational wave spectra from phase transitions},\ }\href@noop {} {\bibfield  {journal} {\bibinfo  {journal} {arXiv preprint arXiv:2407.02580}\ } (\bibinfo {year} {2024})}\BibitemShut {NoStop}%
\bibitem [{\citenamefont {Buen-Abad}\ \emph {et~al.}(2023)\citenamefont {Buen-Abad}, \citenamefont {Chang},\ and\ \citenamefont {Hook}}]{Buen-Abad:2023hex}%
  \BibitemOpen
  \bibfield  {author} {\bibinfo {author} {\bibfnamefont {M.~A.}\ \bibnamefont {Buen-Abad}}, \bibinfo {author} {\bibfnamefont {J.~H.}\ \bibnamefont {Chang}},\ and\ \bibinfo {author} {\bibfnamefont {A.}~\bibnamefont {Hook}},\ }\bibfield  {title} {\bibinfo {title} {{Gravitational wave signatures from reheating}},\ }\href {https://doi.org/10.1103/PhysRevD.108.036006} {\bibfield  {journal} {\bibinfo  {journal} {Phys. Rev. D}\ }\textbf {\bibinfo {volume} {108}},\ \bibinfo {pages} {036006} (\bibinfo {year} {2023})},\ \Eprint {https://arxiv.org/abs/2305.09712} {arXiv:2305.09712 [hep-ph]} \BibitemShut {NoStop}%
\bibitem [{\citenamefont {Schmitz}(2021)}]{Schmitz:2020syl}%
  \BibitemOpen
  \bibfield  {author} {\bibinfo {author} {\bibfnamefont {K.}~\bibnamefont {Schmitz}},\ }\bibfield  {title} {\bibinfo {title} {{New Sensitivity Curves for Gravitational-Wave Signals from Cosmological Phase Transitions}},\ }\href {https://doi.org/10.1007/JHEP01(2021)097} {\bibfield  {journal} {\bibinfo  {journal} {JHEP}\ }\textbf {\bibinfo {volume} {01}},\ \bibinfo {pages} {097}},\ \Eprint {https://arxiv.org/abs/2002.04615} {arXiv:2002.04615 [hep-ph]} \BibitemShut {NoStop}%
\bibitem [{\citenamefont {Batell}\ \emph {et~al.}(2024)\citenamefont {Batell}, \citenamefont {Ghalsasi}, \citenamefont {Low},\ and\ \citenamefont {Rai}}]{Batell:2023wdb}%
  \BibitemOpen
  \bibfield  {author} {\bibinfo {author} {\bibfnamefont {B.}~\bibnamefont {Batell}}, \bibinfo {author} {\bibfnamefont {A.}~\bibnamefont {Ghalsasi}}, \bibinfo {author} {\bibfnamefont {M.}~\bibnamefont {Low}},\ and\ \bibinfo {author} {\bibfnamefont {M.}~\bibnamefont {Rai}},\ }\bibfield  {title} {\bibinfo {title} {{Gravitational Waves from Nnaturalness}},\ }\href {https://doi.org/10.1007/JHEP01(2024)148} {\bibfield  {journal} {\bibinfo  {journal} {JHEP}\ }\textbf {\bibinfo {volume} {01}},\ \bibinfo {pages} {148}},\ \Eprint {https://arxiv.org/abs/2310.06905} {arXiv:2310.06905 [hep-ph]} \BibitemShut {NoStop}%
\bibitem [{\citenamefont {Fedderke}\ \emph {et~al.}(2022)\citenamefont {Fedderke}, \citenamefont {Graham},\ and\ \citenamefont {Rajendran}}]{Fedderke:2021kuy}%
  \BibitemOpen
  \bibfield  {author} {\bibinfo {author} {\bibfnamefont {M.~A.}\ \bibnamefont {Fedderke}}, \bibinfo {author} {\bibfnamefont {P.~W.}\ \bibnamefont {Graham}},\ and\ \bibinfo {author} {\bibfnamefont {S.}~\bibnamefont {Rajendran}},\ }\bibfield  {title} {\bibinfo {title} {{Asteroids for \ensuremath{\mu}Hz gravitational-wave detection}},\ }\href {https://doi.org/10.1103/PhysRevD.105.103018} {\bibfield  {journal} {\bibinfo  {journal} {Phys. Rev. D}\ }\textbf {\bibinfo {volume} {105}},\ \bibinfo {pages} {103018} (\bibinfo {year} {2022})},\ \Eprint {https://arxiv.org/abs/2112.11431} {arXiv:2112.11431 [gr-qc]} \BibitemShut {NoStop}%
\bibitem [{\citenamefont {Bernal}\ \emph {et~al.}(2021)\citenamefont {Bernal}, \citenamefont {Hajkarim},\ and\ \citenamefont {Xu}}]{Bernal:2021yyb}%
  \BibitemOpen
  \bibfield  {author} {\bibinfo {author} {\bibfnamefont {N.}~\bibnamefont {Bernal}}, \bibinfo {author} {\bibfnamefont {F.}~\bibnamefont {Hajkarim}},\ and\ \bibinfo {author} {\bibfnamefont {Y.}~\bibnamefont {Xu}},\ }\bibfield  {title} {\bibinfo {title} {{Axion Dark Matter in the Time of Primordial Black Holes}},\ }\href {https://doi.org/10.1103/PhysRevD.104.075007} {\bibfield  {journal} {\bibinfo  {journal} {Phys. Rev. D}\ }\textbf {\bibinfo {volume} {104}},\ \bibinfo {pages} {075007} (\bibinfo {year} {2021})},\ \Eprint {https://arxiv.org/abs/2107.13575} {arXiv:2107.13575 [hep-ph]} \BibitemShut {NoStop}%
\bibitem [{\citenamefont {Bernal}\ \emph {et~al.}(2022)\citenamefont {Bernal}, \citenamefont {Fong}, \citenamefont {Perez-Gonzalez},\ and\ \citenamefont {Turner}}]{Bernal:2022pue}%
  \BibitemOpen
  \bibfield  {author} {\bibinfo {author} {\bibfnamefont {N.}~\bibnamefont {Bernal}}, \bibinfo {author} {\bibfnamefont {C.~S.}\ \bibnamefont {Fong}}, \bibinfo {author} {\bibfnamefont {Y.~F.}\ \bibnamefont {Perez-Gonzalez}},\ and\ \bibinfo {author} {\bibfnamefont {J.}~\bibnamefont {Turner}},\ }\bibfield  {title} {\bibinfo {title} {{Rescuing high-scale leptogenesis using primordial black holes}},\ }\href {https://doi.org/10.1103/PhysRevD.106.035019} {\bibfield  {journal} {\bibinfo  {journal} {Phys. Rev. D}\ }\textbf {\bibinfo {volume} {106}},\ \bibinfo {pages} {035019} (\bibinfo {year} {2022})},\ \Eprint {https://arxiv.org/abs/2203.08823} {arXiv:2203.08823 [hep-ph]} \BibitemShut {NoStop}%
\bibitem [{\citenamefont {Laurent}\ and\ \citenamefont {Cline}(2022)}]{Laurent:2022jrs}%
  \BibitemOpen
  \bibfield  {author} {\bibinfo {author} {\bibfnamefont {B.}~\bibnamefont {Laurent}}\ and\ \bibinfo {author} {\bibfnamefont {J.~M.}\ \bibnamefont {Cline}},\ }\bibfield  {title} {\bibinfo {title} {{First principles determination of bubble wall velocity}},\ }\href {https://doi.org/10.1103/PhysRevD.106.023501} {\bibfield  {journal} {\bibinfo  {journal} {Phys. Rev. D}\ }\textbf {\bibinfo {volume} {106}},\ \bibinfo {pages} {023501} (\bibinfo {year} {2022})},\ \Eprint {https://arxiv.org/abs/2204.13120} {arXiv:2204.13120 [hep-ph]} \BibitemShut {NoStop}%
\bibitem [{\citenamefont {Basler}\ \emph {et~al.}(2024)\citenamefont {Basler}, \citenamefont {Biermann}, \citenamefont {M{\"u}hlleitner}, \citenamefont {M{\"u}ller}, \citenamefont {Santos},\ and\ \citenamefont {Viana}}]{Basler:2024aaf}%
  \BibitemOpen
  \bibfield  {author} {\bibinfo {author} {\bibfnamefont {P.}~\bibnamefont {Basler}}, \bibinfo {author} {\bibfnamefont {L.}~\bibnamefont {Biermann}}, \bibinfo {author} {\bibfnamefont {M.}~\bibnamefont {M{\"u}hlleitner}}, \bibinfo {author} {\bibfnamefont {J.}~\bibnamefont {M{\"u}ller}}, \bibinfo {author} {\bibfnamefont {R.}~\bibnamefont {Santos}},\ and\ \bibinfo {author} {\bibfnamefont {J.}~\bibnamefont {Viana}},\ }\href@noop {} {\bibinfo {title} {Bsmpt v3: A tool for phase transitions and primordial gravitational waves in extended higgs sectors}} (\bibinfo {year} {2024}),\ \bibinfo {note} {preprint},\ \Eprint {https://arxiv.org/abs/2404.19037} {arXiv:2404.19037 [hep-ph]} \BibitemShut {NoStop}%
\bibitem [{\citenamefont {Azatov}\ \emph {et~al.}(2024{\natexlab{b}})\citenamefont {Azatov}, \citenamefont {Barni}, \citenamefont {Petrossian-Byrne},\ and\ \citenamefont {Vanvlasselaer}}]{Azatov:2023xem}%
  \BibitemOpen
  \bibfield  {author} {\bibinfo {author} {\bibfnamefont {A.}~\bibnamefont {Azatov}}, \bibinfo {author} {\bibfnamefont {G.}~\bibnamefont {Barni}}, \bibinfo {author} {\bibfnamefont {R.}~\bibnamefont {Petrossian-Byrne}},\ and\ \bibinfo {author} {\bibfnamefont {M.}~\bibnamefont {Vanvlasselaer}},\ }\bibfield  {title} {\bibinfo {title} {Quantisation across bubble walls and friction},\ }\href {https://doi.org/10.1007/JHEP05(2024)294} {\bibfield  {journal} {\bibinfo  {journal} {JHEP}\ }\textbf {\bibinfo {volume} {05}},\ \bibinfo {pages} {294}},\ \Eprint {https://arxiv.org/abs/2310.06972} {arXiv:2310.06972 [hep-ph]} \BibitemShut {NoStop}%
\bibitem [{\citenamefont {Hindmarsh}\ \emph {et~al.}(2015)\citenamefont {Hindmarsh}, \citenamefont {Huber}, \citenamefont {Rummukainen},\ and\ \citenamefont {Weir}}]{Hindmarsh:2015qta}%
  \BibitemOpen
  \bibfield  {author} {\bibinfo {author} {\bibfnamefont {M.}~\bibnamefont {Hindmarsh}}, \bibinfo {author} {\bibfnamefont {S.~J.}\ \bibnamefont {Huber}}, \bibinfo {author} {\bibfnamefont {K.}~\bibnamefont {Rummukainen}},\ and\ \bibinfo {author} {\bibfnamefont {D.~J.}\ \bibnamefont {Weir}},\ }\bibfield  {title} {\bibinfo {title} {{Numerical simulations of acoustically generated gravitational waves at a first order phase transition}},\ }\href {https://doi.org/10.1103/PhysRevD.92.123009} {\bibfield  {journal} {\bibinfo  {journal} {Phys. Rev. D}\ }\textbf {\bibinfo {volume} {92}},\ \bibinfo {pages} {123009} (\bibinfo {year} {2015})},\ \Eprint {https://arxiv.org/abs/1504.03291} {arXiv:1504.03291 [astro-ph.CO]} \BibitemShut {NoStop}%
\bibitem [{\citenamefont {Guo}\ \emph {et~al.}(2021)\citenamefont {Guo}, \citenamefont {Sinha}, \citenamefont {Vagie},\ and\ \citenamefont {White}}]{Guo:2020grp}%
  \BibitemOpen
  \bibfield  {author} {\bibinfo {author} {\bibfnamefont {H.-K.}\ \bibnamefont {Guo}}, \bibinfo {author} {\bibfnamefont {K.}~\bibnamefont {Sinha}}, \bibinfo {author} {\bibfnamefont {D.}~\bibnamefont {Vagie}},\ and\ \bibinfo {author} {\bibfnamefont {G.}~\bibnamefont {White}},\ }\bibfield  {title} {\bibinfo {title} {{Phase Transitions in an Expanding Universe: Stochastic Gravitational Waves in Standard and Non-Standard Histories}},\ }\href {https://doi.org/10.1088/1475-7516/2021/01/001} {\bibfield  {journal} {\bibinfo  {journal} {JCAP}\ }\textbf {\bibinfo {volume} {01}},\ \bibinfo {pages} {001}},\ \Eprint {https://arxiv.org/abs/2007.08537} {arXiv:2007.08537 [hep-ph]} \BibitemShut {NoStop}%
\bibitem [{\citenamefont {Hindmarsh}\ \emph {et~al.}(2017)\citenamefont {Hindmarsh}, \citenamefont {Huber}, \citenamefont {Rummukainen},\ and\ \citenamefont {Weir}}]{Hindmarsh:2017gnf}%
  \BibitemOpen
  \bibfield  {author} {\bibinfo {author} {\bibfnamefont {M.}~\bibnamefont {Hindmarsh}}, \bibinfo {author} {\bibfnamefont {S.~J.}\ \bibnamefont {Huber}}, \bibinfo {author} {\bibfnamefont {K.}~\bibnamefont {Rummukainen}},\ and\ \bibinfo {author} {\bibfnamefont {D.~J.}\ \bibnamefont {Weir}},\ }\bibfield  {title} {\bibinfo {title} {{Shape of the acoustic gravitational wave power spectrum from a first order phase transition}},\ }\href {https://doi.org/10.1103/PhysRevD.96.103520} {\bibfield  {journal} {\bibinfo  {journal} {Phys. Rev. D}\ }\textbf {\bibinfo {volume} {96}},\ \bibinfo {pages} {103520} (\bibinfo {year} {2017})},\ \bibinfo {note} {[Erratum: Phys.Rev.D 101, 089902 (2020)]},\ \Eprint {https://arxiv.org/abs/1704.05871} {arXiv:1704.05871 [astro-ph.CO]} \BibitemShut {NoStop}%
\end{thebibliography}%

\end{document}